\documentclass[article, reprint, superscriptaddress, prc, showkeys, showpacs,14pt]{revtex4}

\usepackage{amsmath} 
\usepackage{braket}
\usepackage{bm}
\usepackage{graphicx} 
\usepackage{color} 
\usepackage{subfigure} 
\usepackage{hyperref} 

\begin{document}
\title{Investigating Possible Existence of Hyper-Heavy Nuclei \\ in Neutron Star Environment} 

\author{M. Veselsk\'y}  \email{Martin.Veselsky@utef.cvut.cz} 
\affiliation{Institute of Experimental and Applied Physics - Czech Technical University in Prague}
\author{V. Petousis} \email{vlasios.petousis@cern.ch}
\affiliation{Institute of Experimental and Applied Physics - Czech Technical University in Prague}
\author{Ch. C. Moustakidis} 
\affiliation{Aristotle University of Thessaloniki, Thessaloniki, Greece}
\author{G. A. Souliotis}
\affiliation{Laboratory of Physical Chemistry, Department of Chemistry, National and Kapodistrian University of Athens, Athens, Greece}
\author{A. Bonasera}
\affiliation{Cyclotron Institute, Texas A\&M University, College Station, Texas, USA}
\affiliation{Laboratori Nazionali del Sud, INFN, Catania, Italy}

\date{\today}

\begin{abstract}
The synthesis of hyper-heavy elements is investigated under conditions simulating neutron star environment. The Constrained Molecular Dynamics (CoMD) approach is used to simulate low energy collisions of extremely n-rich nuclei. A new type of the fusion barrier due to a "neutron wind" is observed when the effect of neutron star environment (screening of Coulomb interaction) is introduced implicitly. When introducing also a background of surrounding nuclei, the nuclear fusion becomes possible down to temperatures of 10$^{8}$ K and synthesis of extremely heavy and n-rich nuclei appears feasible. A possible existence of hyper-heavy nuclei in a neutron star environment could provide a mechanism of extra coherent neutrino scattering or an additional mechanism, resulting in X-ray burst or a gravitational wave signal and, thus, becoming another crucial process adding new information to the suggested models on neutron star evolution. 
\end{abstract}


\maketitle

\section{Introduction}
We know today that the heaviest elements with Z = 114 - 118 were produced  in hot fusion reactions with emission of 3-4 neutrons using $^{48}$Ca beams and heavy actinide targets between uranium and californium \cite{SHE}. The main obstacle for production of such nuclei is a fusion hindrance caused by competition of  a fusion process with an alternative process called quasi-fission. Quasi-fission occurs, when instead of fusion, the system forms an elongated shape and evolves towards the scission point. It is usually considered that the process of quasi-fission is governed by the complex dynamics of the projectile-target system, where the repulsive Coulomb force acts as a driving force towards separation into two fragments and thus constitutes the principal obstacle to the production of heavier nuclei in laboratory conditions. While such an obstacle can be hardly circumvented in the physical laboratory, there exist an environment, namely neutron stars, where in electrically neutral neutron matter the effect of repulsive Coulomb force can practically vanish. Under conditions of overall electrical neutrality, also the fusion dynamics may appear different. It can be assumed that a heavy nucleus in such environment with density comparable to saturation density will be surrounded by a corresponding electron cloud (not an atomic envelope) which will make nuclei look as neutral even at mutual distances of several $fm$. Such collisions even at low energy, can in principle be simulated, potentially leading to hyper-heavy nuclei in the interior of a neutron star.

Thus, exotic configurations  with $Z>>120$ are hardly relevant to terrestrial (laboratory) experiments, but they can exist in neutron-rich environments included the atmosphere of stellar explosions and the compositions of neutron stars~\cite{Giuliani-2019}. The creation and stability of superheavy and hyperheavy nuclei in the form of bubbles or semi-bubbles (for a large array  of proton and neutron numbers, $120 \leq Z \leq 340$ and $300 \leq A \leq 1000$) ) have been discussed also in~\cite{Decharge-1999}.  

In the seminal paper of Baym {\it et al.}~\cite{Baym-1971}, the authors found that, concerning the structure of  the cold  inner crust, for baryon densities $n_b> 0.006 \ {\rm fm}^{-3}$ ($\rho_{0}$/30) up to the crust-core transition $n_{\rm tr}=0.11 \ {\rm fm}^{-3} $ (2 $\rho_0$/3 where $\rho_0$ is the typical value of the nuclear saturation density), the values of atomic and mass number are within the range $Z=50-200$ and $A=186-2500$ respectively. Other additional studies~\cite{Steiner-2008,Baldo-2007,Pearson-2012} have confirmed the results of~\cite{Baym-1971}.      

An extensive discussion about the neutron star structure and properties both for cold and warm environment has been provided in~\cite{Haensel-2007,Bertulani-2012}. In particular, in Chapter 14 of~\cite{Bertulani-2012}, Page and Reddy discussed the composition of the inner crust in the framework of mean field nuclear models, based on the Skyrme interaction. They considered the dependence of  $A_{\rm cell}$,  $N$, $Z$, $A=N+Z$, and $N_{\rm out}$ on the density in the inner crust and for three cases: (a) cold catalyzed matter, (b) matter in accreting neutron stars, where reactions including pycno-nuclear fusion processes of $^{56}$Fe at $\rho\simeq 10^9{\rm g cm^{-3}}$ to higher densities and (c) matter in accreting neutron star, where is assumed not occur pycno-nuclear fusion for $Z>4$. They show that the total number of nucleons per unit cell (containing one nucleus), which is given by $A_{\rm cell}=N+Z+N_{\rm out}$ ($N_{\rm out}$ is the number of neutrons with single particle energy $>0$), in the cell reaches in all the mentioned cases very high values ($A_{\rm cell}>>100$). Moreover, in a recent study~\cite{Grams-2021}, the authors, using the more elaborated $\chi_{\rm EFT}$ Hamiltonian and the SLy4 phenomenological force, calculated the composition of the neutron star crust. In particular they predicted the cluster mass number $A_{\rm cl}$ and proton number $Z_{\rm cl}$ through the crust, which correspond to the masses and proton numbers associated to the nuclear cluster only (excluding the neutron fluid contribution). They also found very high values for $A_{\rm cl}$ and $Z_{\rm cl}$ close to the crust-core interface (baryon density $\simeq 0.1 $ fm$^{-3}$).

The decompression of the inner crust, following the dynamically ejected mass after neutron star merger,  has been discussed in~\cite{Goriely-2011}. In their calculations, nuclei up to $Z=100$ were involved. All fusion reactions on light elements that play a role on nuclear statistical equilibrium are included. In addition, fission and $\beta$-decays are
also included, for example  neutron-induced fission, spontaneous fission, $\beta$-delayed fission, photofission, as well as $\beta$-delayed neutron emission. 
The authors concluded that matter from the inner crust of the coalescing NSs, which by far dominates the ejecta, produces an r-abundance distribution very similar to the
solar one for nuclei with $A>140$~\cite{Goriely-2011}.    
    
In a notable study \cite{Shen-2011}, Shen et al constructed  a large class of hot equations of state for use in astrophysical simulations of supernovae and neutron star mergers. In particular, they discussed the structure of inner crust in a relativistic mean field model with spherical Wigner-Seitz approximation, at finite temperature and proton fraction over a wide range of densities. They found that the average mass number $A$ can be as large as $3000$ at high density before the final transition to uniform matter. At higher temperatures, as shown~\cite{Shen-2011}, $A$ grows more rapidly to several thousand in narrower ranges of density. Moreover, the average proton number $Z$ of heavy nuclei (versus baryon density at different temperatures and proton number), may be even higher than $1000$.

According to the above discussion, theoretical studies suggest that the existence of nuclei with extremely high values of $Z$ and $A$ are possible. Consequently  there is an open issue to investigate the possibility of fusion process to take place in the environment of hot or/and cold neutron star crust. The instability of these super-heavy nuclei is another  issue worth to be studied. In any case,  useful insights  can be gained in the  effort to investigate the possibility of creation of such nuclei in the interior of a neutron star.   

The existence of finite nuclei in neutron star environment is crucial for the solution of the problem of their cooling time, since the reactions with neutrinos can limit its cooling rate \cite{171010441}. Ways of producing nuclei in neutron stars were summarized in \cite{180303818}, where a reaction network in a neutron star crust is described. 
Here we investigate the conditions under which eventually hyper-heavy nuclei can be produced by fusion of very n-rich heavy nuclei using Constrained Molecular Dynamics (CoMD) simulations. This might be a possible way to circumvent the limitations observed in the synthesis of hyper-heavy elements. 

Summarizing the existence of extremely n-rich nuclei, with N/Z ratios much higher than those experimentally observed nuclei, can be expected in a neutron star environment, where they will be surrounded by dilute neutron matter and their stability will depend on the balance of inward and outward flow of neutrons. Such nuclei can be possibly considered as a special case of nuclear pasta and they can, in principle, undergo fusion and form extremely heavy nuclei. Existence of such hyper-heavy nuclei was investigated recently in \cite{180406395}. In this work, we investigate conditions of fusion in reactions leading to production of hyper-heavy nuclei up to Z = 120 and beyond. 

We  note that using the Boltzmann-Uhling-Uhlenbeck equation (BUU)~\cite{BUUSHE}, it was possible to set a rather strict constraint on the incompressibility of the equation of state of nuclear matte,r $K_0 = 254~MeV$, with softer density dependence of symmetry energy with $\gamma = 0.5 - 1.0$. 
The same parameters were obtained also when the CoMD transport code was used \cite{CoMDSHE}. We thus consider feasible to use such an equation of state also for simulations of nucleus-nucleus collisions in a neutron star environment. We performed such calculations in the present work. 

This article is structured as follows. In section 2 we present the CoMD theoretical framework, in section 3 we explain and present the results of our simulations without and with nucleonic medium and we close with our conclusions and the relevant references.

\section{Theoretical Framework}
The Constrained Molecular Dynamics (CoMD) model \cite{CoMD} has been used recently to describe the dynamics of low-energy phenomena. Studies incorporating reactions near the Fermi energy \cite{George18,George19,George20,George21} and the fission of heavy nuclei \cite{George22,George23}, have been performed. The CoMD model is based on a two-body effective interaction with Skyrme potential characteristics which are written as:  

\begin{equation}
V_{vol} = \frac{T_{0}}{\rho_{0}}\delta(\pmb{r_{i}-r_{j}})
\end{equation}
\begin{equation}
V_{3} = \frac{2T_{3}\rho^{\sigma-1}}{(\sigma+1)\rho_{0}^{\sigma}}\delta(\pmb{r_{i}-r_{j}})
\end{equation}
\begin{equation}
V_{sym} = \frac{\alpha_{sym}}{\rho_{0}}\delta(\pmb{r_{i}-r_{j}})(2\delta_{\tau_{i},\tau_{j}}-1)
\end{equation}
\begin{equation}
V_{sur} = \frac{C_{s}}{\rho_{0}}\nabla^{2}_{\braket{\pmb{r}_{i}}}\delta(\pmb{r_{i}-r_{j}})
\end{equation}
\begin{equation}
V_{coul} = \frac{e^{2}}{\left \lvert \lvert {\pmb{r_{i}-r_{j}}} \right \rvert \rvert}
\end{equation}

In the above set of potentials, $V_{vol}$ corresponds to the volume term, $V_{3}$ to the three-body term, $V_{sym}$ to the symmetry term and $V_{sur}$ to the surface term, while $V_{coul}$ describes the Coulomb repulsion of the interacting protons. in this work, the parameters $\rho_{0}$, $\alpha_{sym}$ and $C_{s}$ are the saturation density, the symmetry and surface parameter, respectively. Their values are fixed namely to ${\rho_{0}=0.16~fm^{-3}}$, ${\alpha_{sym}=32~MeV}$ and ${C_{s}/\rho_{0}=0}$. The constants  $T_{0}$,  $T_{3}$ and  $\sigma$ are related to the compressibility of Nuclear Matter (NM) and are determined by the enforcement of the saturation condition of NM with the use of the effective interaction involving eq.~(1) and  eq.~(2). The one-body wave-functions used in the model are Gaussian wave-packets in the Wigner representation having the form:

\begin{equation}
f(\pmb{r,p}) = \frac{1}{({2\pi\sigma_{r}\sigma_{p}})^{3}}e^{-\frac{(\pmb{r}-\braket{\pmb{r_{j}}})^{2}}{2\sigma_{r}^{2}}}e^{-\frac{(\pmb{p}-\braket{\pmb{p}_{j}})^{2}}{2\sigma_{p}^{2}}}
\end{equation}

The  ${\sigma_{r}}$ and ${\sigma_{p}}$ are the widths, while $\braket{\pmb{r_{i}}}$ and $\braket{\pmb{p_{i}}}$ are the time dependent centroids of the wave-packet of nucleon ${\pmb{i}}$ in phase space. In the calculations presented below, ${\sigma_{r}=1.065~fm}$ and ${\sigma_{p}\sigma_{r}=\hbar/2}$.

\section{Simulations}
Simulations of reactions potentially leading to hyper-heavy nuclei in the neutron star interior started a few years ago \cite{BUUSHE} using BUU, without Coulomb interaction assuming the presence of electrons and thus charge neutrality. We assumed neutron-rich nuclei with N/Z = 4, $^{140}Ni+^{460}U$ in particular. This reaction analogous to reaction $^{64}Ni+^{238}U$, with the highest proton number among the reactions investigated, focusing on reactions, leading to experimentally observed superheavy nuclei. The choice of N/Z = 4 corresponds to expected proton fractions in a neutron star environment. Such BUU simulations observed no fusion barrier except for a repulsive "neutron wind" caused by emission of nucleons, which can prevent contact of nuclei in the otherwise empty space and thus fusion. This effect appears stronger for stiff symmetry energy. It is assumed that n-rich nuclei exist in the inner crust of proto-neutron stars, and their fusion (and eventually formation of hyper-heavy nuclei), can affect cooling due to coherent neutrino scattering. The reactions of nuclei in an inhomogeneous medium \cite{InnercrustPasta} can be strongly influenced by proton fraction and distribution of electrons assuring overall neutrality. There is no a priori reason why electrons (electron fluid) could not enter the nucleus, especially since the calculated Debye-H\"{u}ckel length amounts to a few $fm$. Possible distribution of charge in nuclei can thus remind a so-called plasma mirror on the surface of nuclei, with excess of electrons outwards and excess of protons inside. In this way, the Coulomb interaction will act only at close distance of few $fm$ as given by the Debye-H\"{u}ckel length: 

\begin{equation}
\lambda_{D}= \sqrt{\frac{\epsilon_0}{n_{e} {q_e^{2}}/{T_{e}} + {n_{p}} {q_p^{2}} {/T_{p}}}}
\end{equation}

The Debye-H\"{u}ckel length is determined by densities ($n_{p,e}$) and temperatures ($T_{p,e}$), of both protons and electrons. The simulations have control of the density and temperature of protons, while the density of electron fluid can be assumed equal to the density of protons, even if local deviations are possible, especially around the surface and in the interior of nuclei. The temperature of electron fluid (and thus the localization of electrons) can be considerably higher than the temperature of protons. Thus we choose the value of Debye-H\"{u}ckel length by the density and assumed temperature of proton gas within the nucleonic background, the latter taken equal to the beam energy. 

The Debye-H\"{u}ckel formula used in this work assumes classical protons. This approximation can be justified for lowest values of density of the proton background, where assumed temperature of around 1 MeV it is comparable to Fermi energy. Even for largest assumed proton densities, the Debye-Huckel formula yields values of proton screening length comparable to the Thomas-Fermi formula assuming zero temperature. The approximation used for proton screening thus appears as reasonable. The impact parameter is chosen between 0 and $1fm$, as in previous simulations leading to the synthesis of superheavy nuclei. However for energies comparable to temperatures below a few MeV, 
all the possible reactions proceed within an s-wave, and thus the fusion probabilities will be the same. 

\subsection{Without nucleonic medium}
Recently the CoMD code was modified to include Coulomb screening via an interaction cutoff at a distance corresponding to the Debye-H\"{u}ckel length.
The simulations were performed with the CoMD code, using the EoS with $K_0 = 254~MeV$, which was found to be optimal for the description of reactions leading to the synthesis of superheavy nuclei in the laboratory \cite{CoMDSHE}. The time step was $1~fm/c$ and the charge screening effect of surrounding nuclear medium in the neutron star was simulated by implementing a cutoff, limiting the range of the Coulomb interaction to $2~fm$. This corresponds approximately to the density of surrounding nucleonic medium amounting to ${\rho_{0}/5}$. Fusion barrier was observed between $1.25~MeV$ and $1.5~MeV/nucleon$ (see Fig.1 and Fig.2) due to a "neutron wind" even with Coulomb interaction cutoff of $2~fm$. The turning point is at distances exceeding the range of such screened Coulomb interaction.   

Long term survival up to $30000~fm/c$ was also investigated, using again $^{140}Ni+^{460}U$ at very low kinetic energy (a value $6~keV$ was chosen and the Coulomb interaction cutoff was set at $1fm$). A long term stability was observed for both soft and stiff symmetry energy, even if for stiff symmetry energy nuclei are more unstable against emission of nucleons, causing the observed "neutron wind", which is stronger than in the case of soft symmetry energy. Such behavior is observed up to a Coulomb interaction cutoff of $10fm$. Above such value, Coulomb interaction starts to dominate and nuclei become unstable against fission or fragmentation into more parts.

\begin{figure}[ht]
\begin{minipage}{\textwidth}
\centering
\fbox{\includegraphics[width=8.6cm]{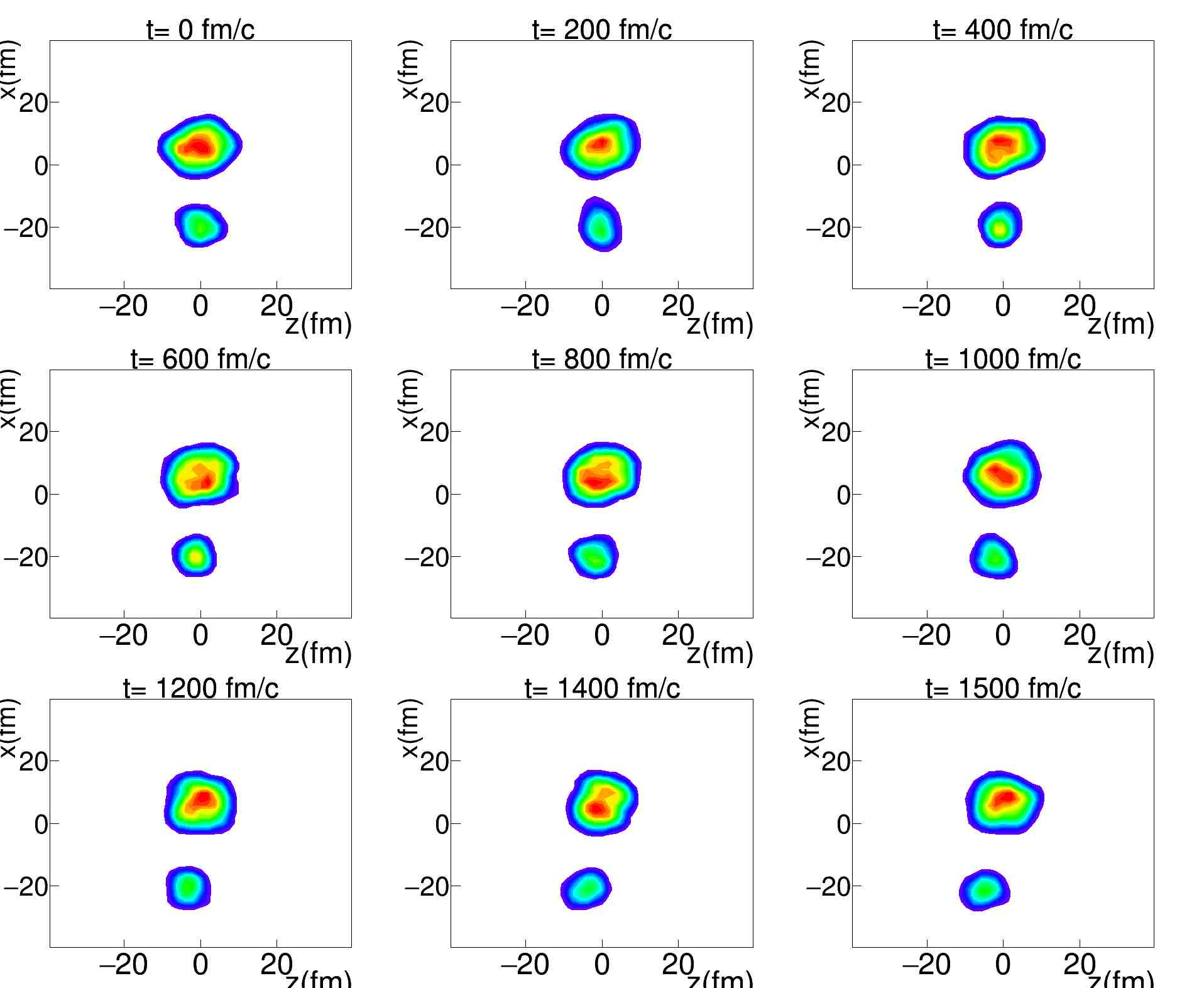}} 
\fbox{\includegraphics[width=8.6cm]{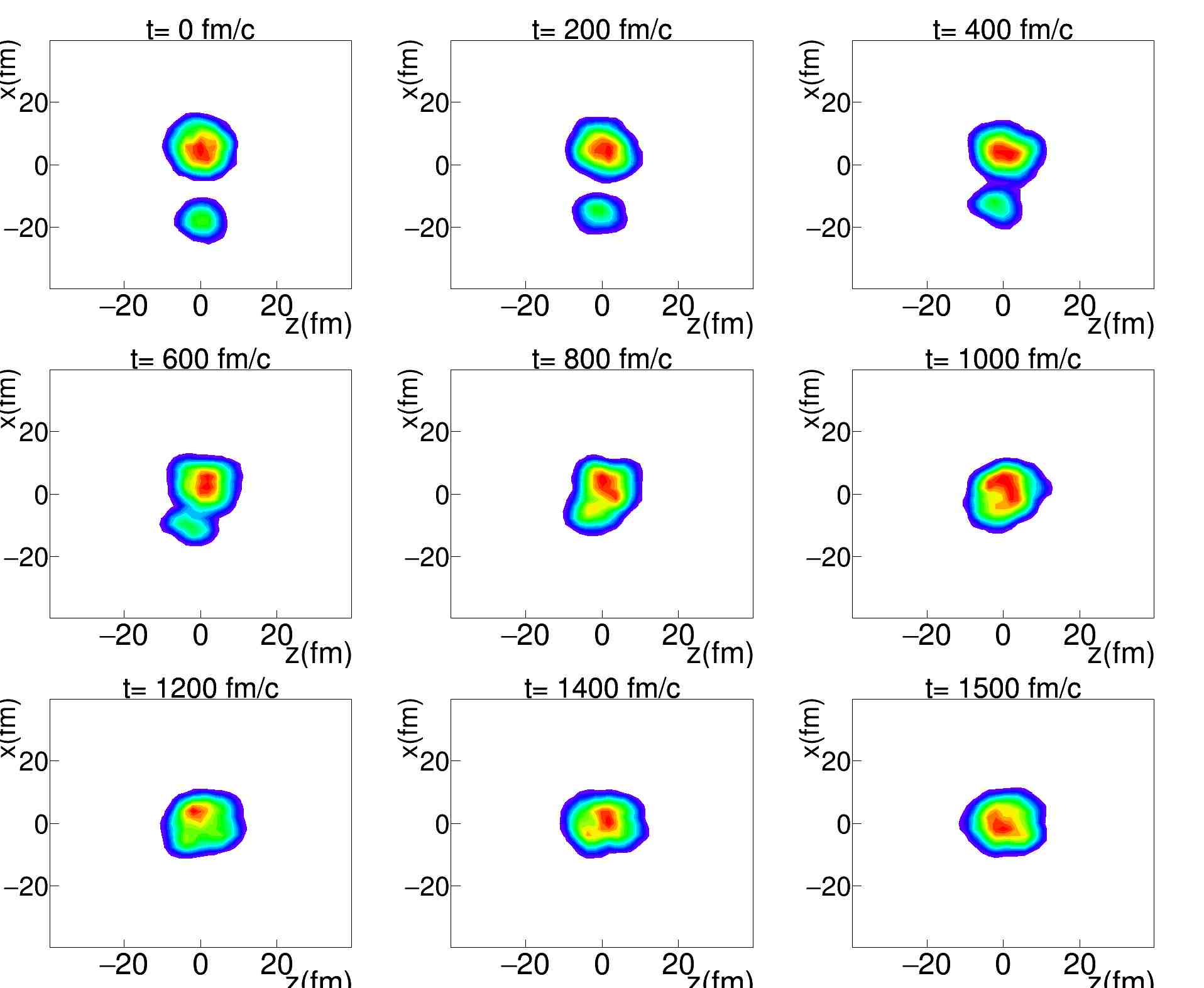}} 
\end{minipage}
\caption{Typical evolution of nucleonic density for central collision $^{140}$Ni+$^{460}$U calculated using the CoMD code at beam energy $1.25~MeV/nucleon$ (Left panel) and $1.5~MeV/nucleon$ (Right panel). A Coulomb interaction cutoff 2 $fm$, incompressibility $K_0 = 254~MeV$ and a soft density dependence of symmetry energy were used. Scattering due to a "neutron wind" dominates at lower beam energy (Left panel), while fusion dominates at higher beam energy (Right panel).}
\end{figure}

\begin{figure}[ht]
\begin{minipage}{\textwidth}
\centering
\fbox{\includegraphics[width=8.6cm]{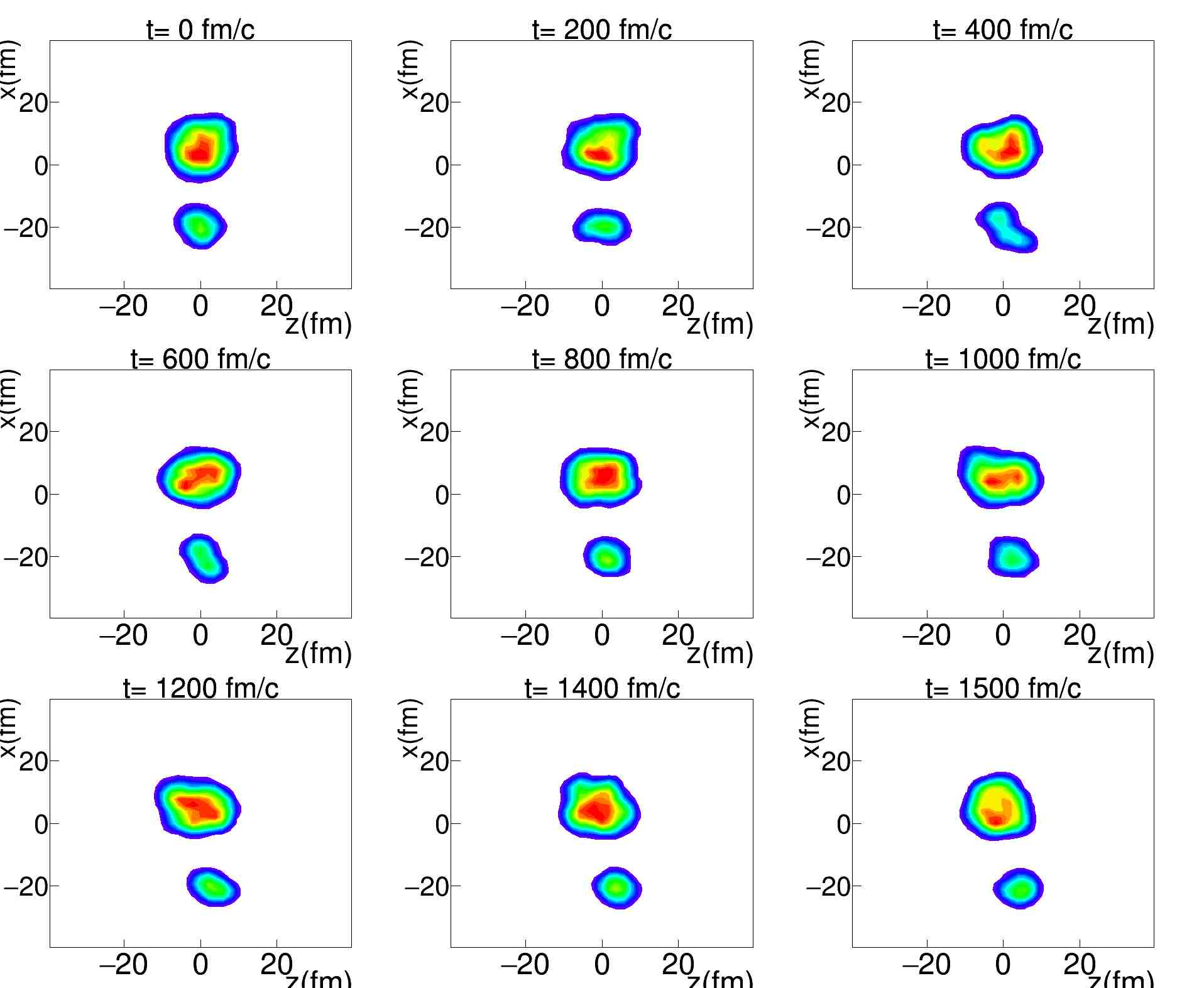}} 
\fbox{\includegraphics[width=8.6cm]{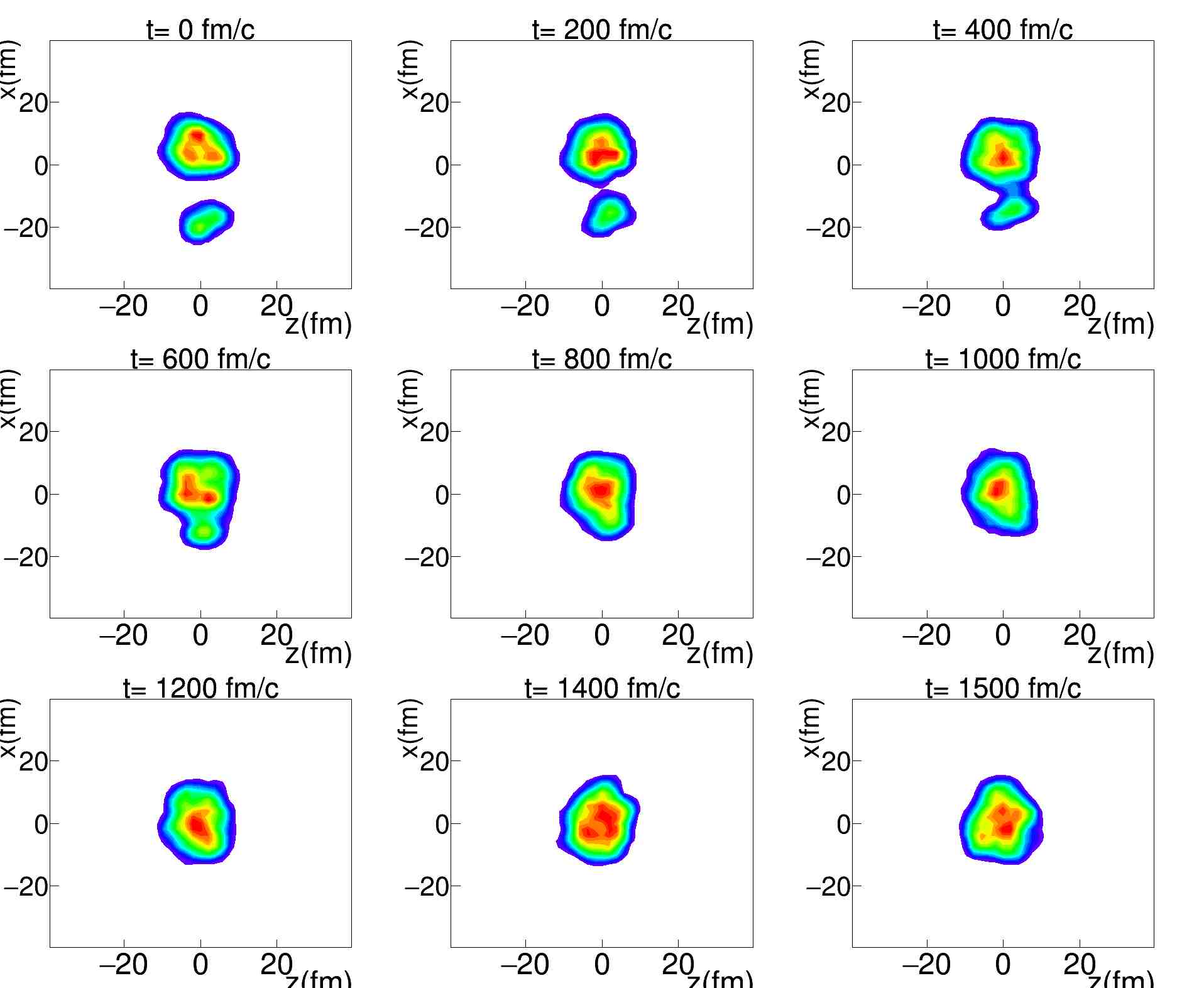}} 
\end{minipage}
\caption{Typical evolution of nucleonic density for central collision $^{140}$Ni+$^{460}$U calculated using the CoMD code at beam energy $1.25~MeV/nucleon$ (Left panel) and $1.5~MeV/nucleon$ (Right panel). A Coulomb interaction cutoff $2fm$, incompressibility $K_0 = 254~MeV$ and a stiff density dependence of symmetry energy were used. Scattering due to a "neutron wind" dominates at lower beam energy (Left panel), while fusion dominates at higher beam energy (Right panel).}
\end{figure}

\subsection{With nucleonic medium}
To simulate conditions inside neutron stars, CoMD was further modified to introduce a low density nucleonic background by placing 2000 nucleons inside a box with periodic boundary conditions. The size of the box was initially set to 50x50x50${~fm^{3}}$, corresponding to approximately one tenth of saturation density. The cutoff of Coulomb interaction was set correspondingly to the Debye-H\"{u}ckel length at given density and proton fraction of nuclear medium (10\%).
The simulations lead to the confirmation that the "neutron wind" observed previously was unphysical and practically there is no fusion barrier down to beam energy corresponding to typical neutron star temperature ${10^{8} - 10^{9}}$ K (${10-100~keV}$). An interesting new observation is that besides fusion, there exists another mechanism of growth of nuclei,  since for soft ($\gamma=0.5$, Fig.3) and intermediate ($\gamma=0.75$, Fig.4) symmetry energy the resulting compound nucleus captures and incorporates nearly all nucleons of the background in the box. This signals an instability which could lead to a significant presence of very heavy nuclei in a neutron star interior, especially prior to cooling down to equilibration temperature where this distribution of nuclear sizes will follow a statistical law. For stiff symmetry energy ($\gamma=1.0$, Fig.5), such effect is much weaker if any at all. Thus the symmetry energy can play an additional role in the structure of proto-neutron stars and their cooling rate.

\subsection{Final simulations}

Besides a background density of one tenth of the saturation density, lower and higher background densities were explored, with the range of Coulomb interaction adjusted to thecorresponding Debye-H\"{u}ckel length. The effect of a collapse of background nucleons into compound nucleus is preserved also at background density one fifth of saturation density, while it disappeares at lower background densities. Concerning the nature of such collapse, it is most probably connected to the first order phase transition in nuclear matter, where the density dependence of symmetry energy controls the extent of spinodal region in the isovector plane.
In order to estimate a possible effect of such reactions on the structure and properties of a neutron star, the simulations were run for longer time up to $25000~fm/c$ ($10^{-20}~s$). If the formed compound nucleus survives for such time, it will have enough time to travel to a neighboring cell, where analogous reaction can occur, and cause further reaction with a neighboring compound nucleus, thus forming even heavier system. Such scenario is supported by the work \cite{Shen-2011}, where fractions of heavy nuclei up to 70\% are predicted in the density range $\rho_0$/5 -- $\rho_0$/50. The simulations in Figs.~3,4,5 were performed using kinetic energy of $10~keV$ in the c.m. frame, but similar results were obtained also for beam energies of few $MeV$, corresponding to analogous temperatures. This can be naturally expected  when considering that the total energy of the system remains essentially the same. 
It appears that a temperature above $1~MeV$ might be sufficient to provide reaction products with sufficient mobility to support the fusion reaction cascade and a formation of a  heavier system. 
Such temperatures can be explained not only on proto-neutron stars \cite{Kumar, Bz24, Bz55}, but also in cold stable neutron stars \cite{Wei}. The results of simulations are shown in Figs.~3 and 5, for soft, medium and stiff parameterizations of the density dependence of symmetry energy, respectively.    

On the time scale up to $25000~fm/c$ for the soft and medium symmetry energy parameterization, the resulting nuclei appear to dissolve in the nucleon bath with density above ${\rho_{0}/30}$, and even sooner with increasing density. For lower densities the resulting nuclei appear to survive long enough to reach neighboring cells and collide with their equivalent nuclei there. For a stiff parameterization of the symmetry energy, the behavior is similar at high densities, however, at lowest density, onset of fission can be observed, thus reverting the possible fusion reaction cascade. Still, a maximum of lifetime can be observed between densities ${\rho_{0}/30}$ and ${\rho_{0}/50}$, where a fusion reaction cascade may be supported. Similar maximum of lifetime could be possibly observed for softer symmetry energy parameterizations at longer times.
The above simulations were performed at proton fraction of 10\%. Increasing or decreasing the proton fraction, will mean corresponding decreasing or increasing the Debye-H\"{u}ckel length by the square root of the relative increase (decrease) of the proton fraction. This variation can affect the simulations only moderately within the expected proton fractions between a few percent up to 20\%, possibly shifting the optimum density slightly below or above, respectively.  
   
\begin{figure}[h]
\begin{minipage}{\textwidth}
\centering
\fbox{\includegraphics[width=8.6cm]{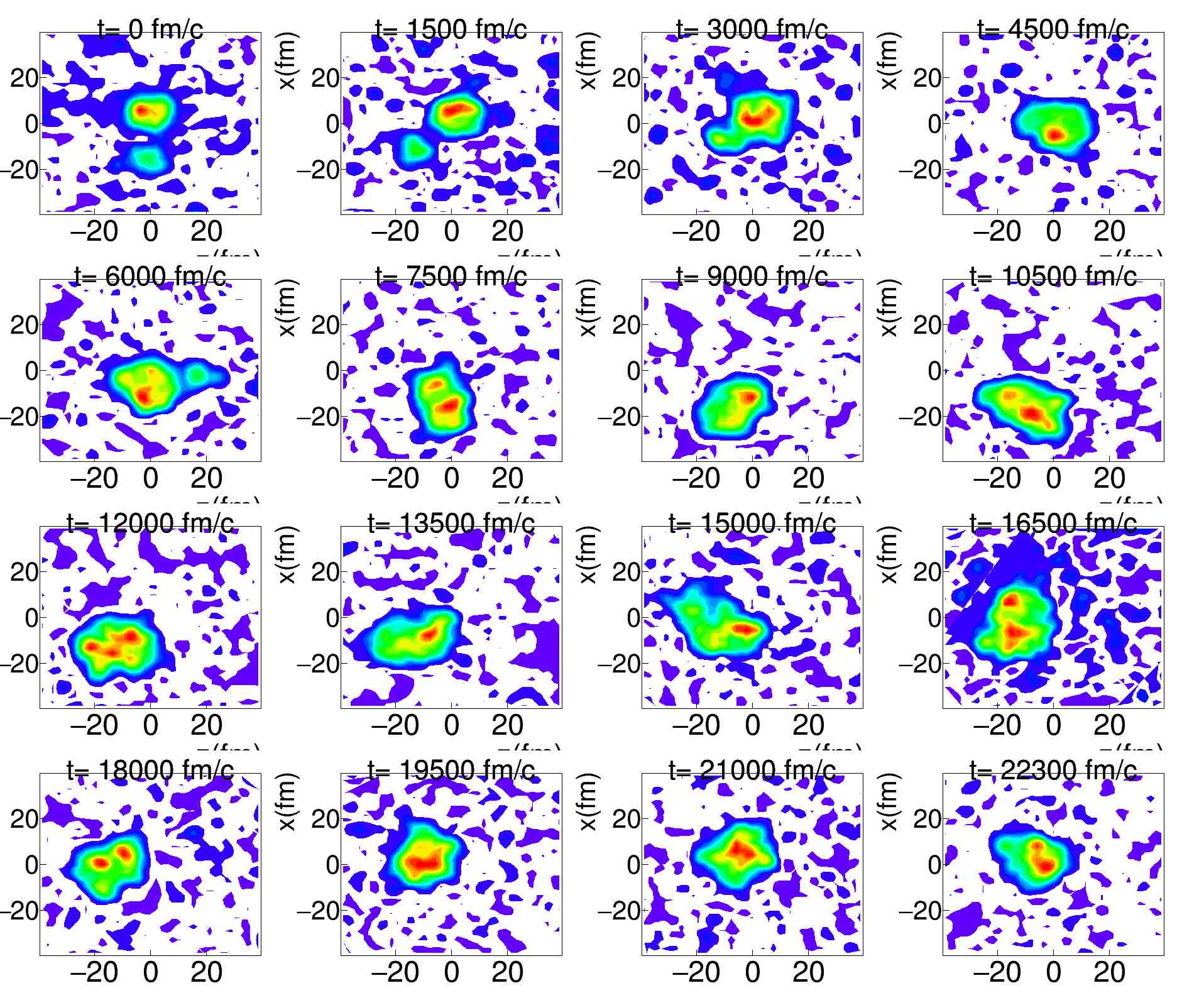}} 
\fbox{\includegraphics[width=8.6cm]{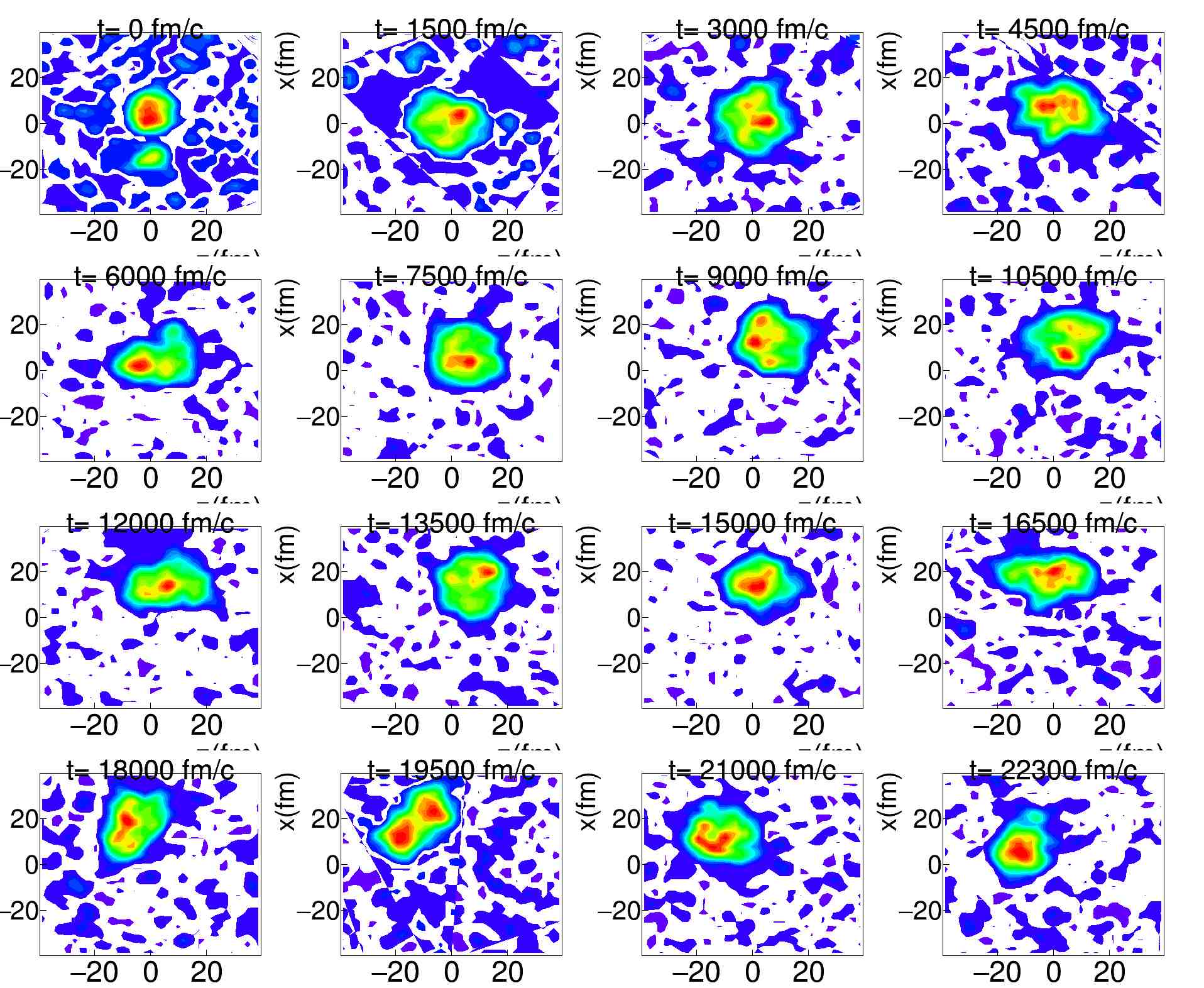}} 
\fbox{\includegraphics[width=8.6cm]{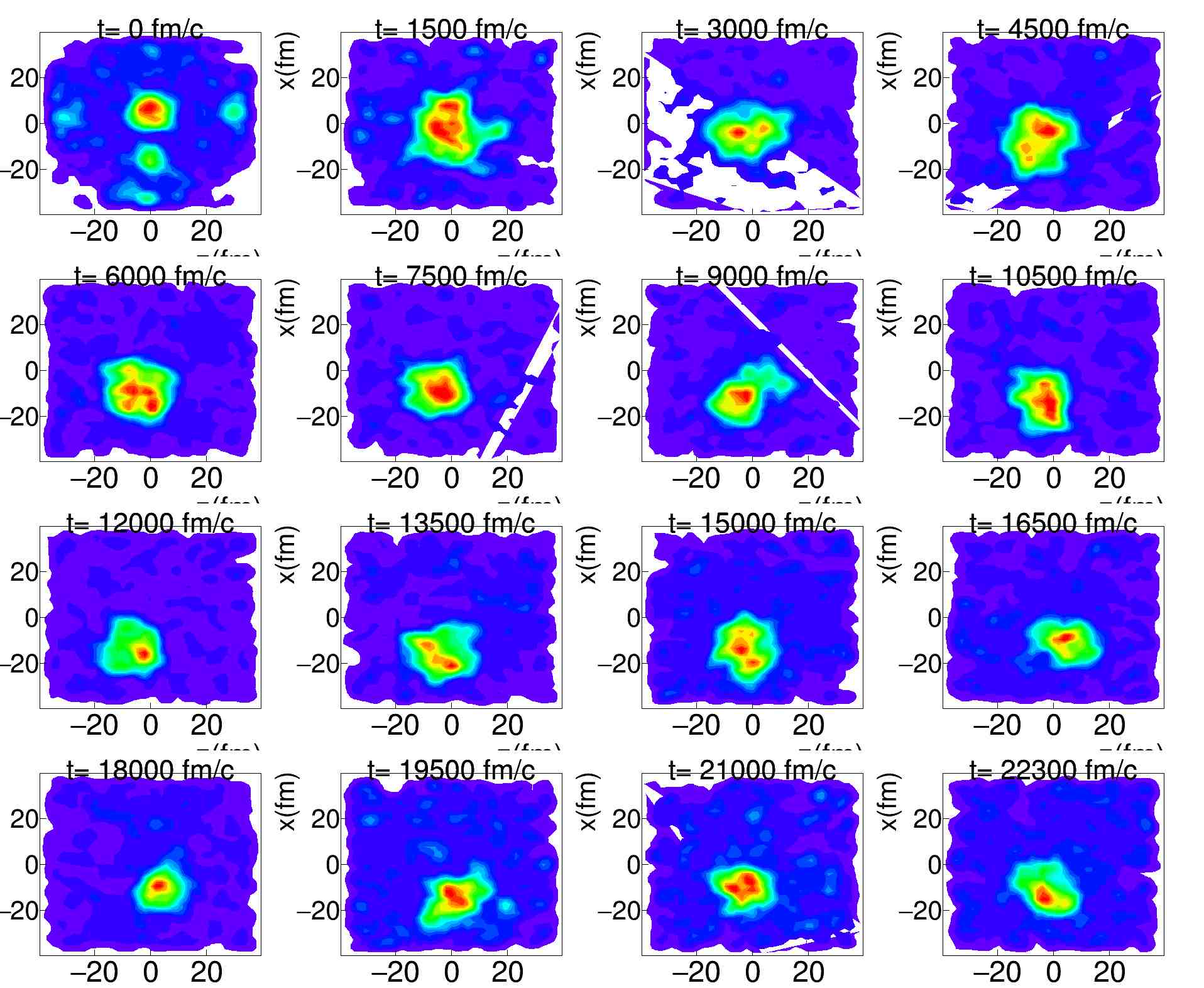}} 
\fbox{\includegraphics[width=8.6cm]{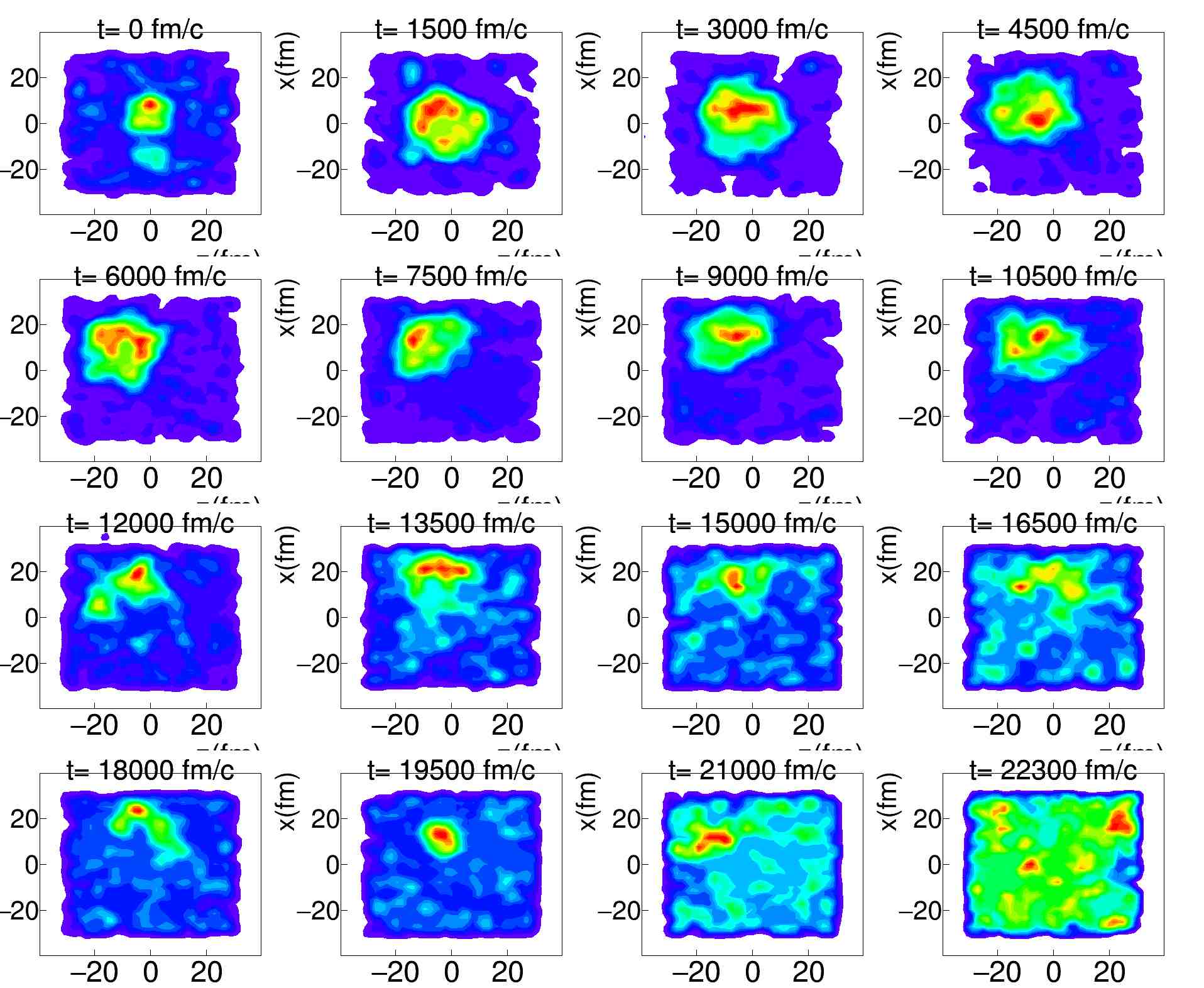}} 
\fbox{\includegraphics[width=8.6cm]{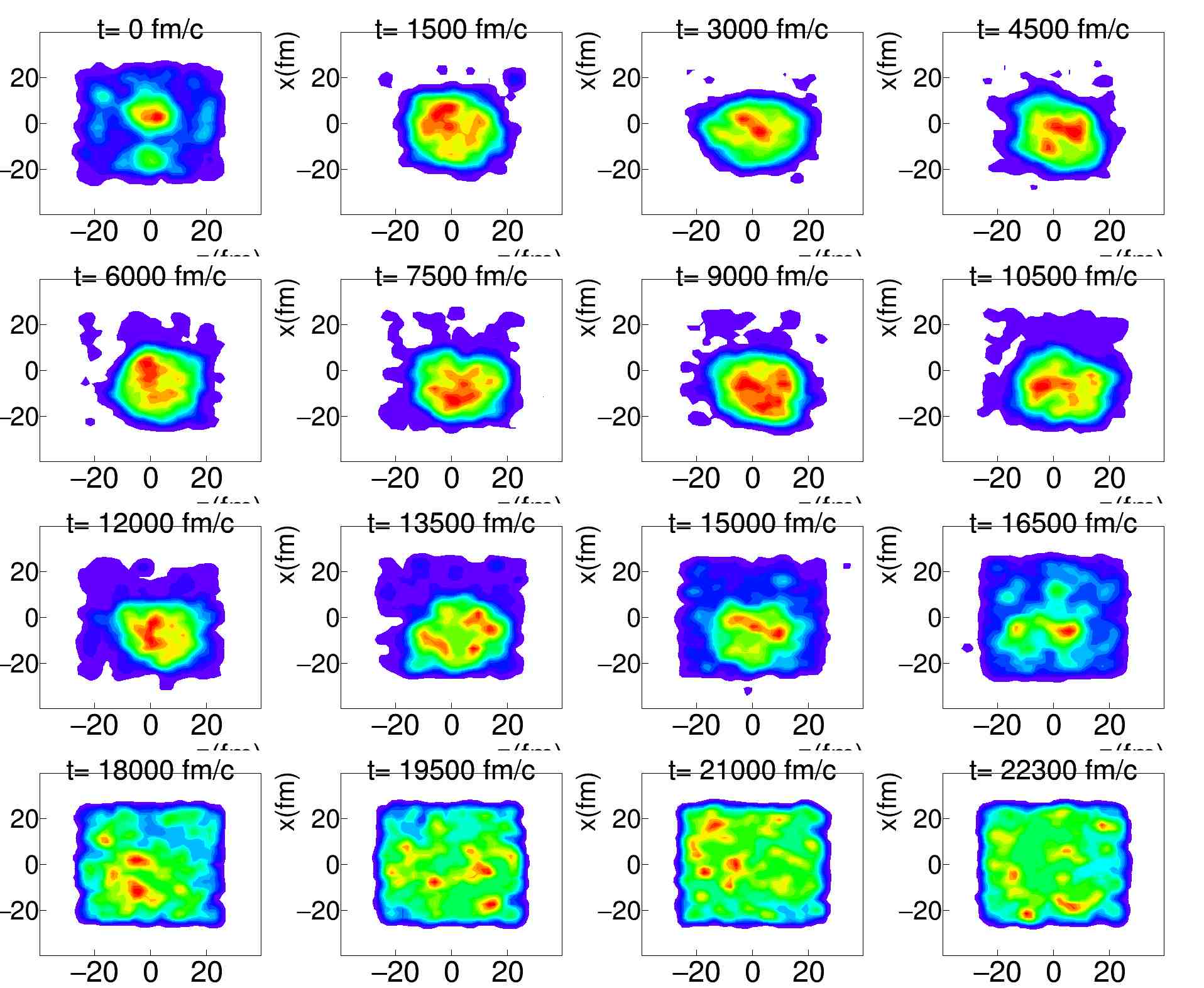}} 
\fbox{\includegraphics[width=8.6cm]{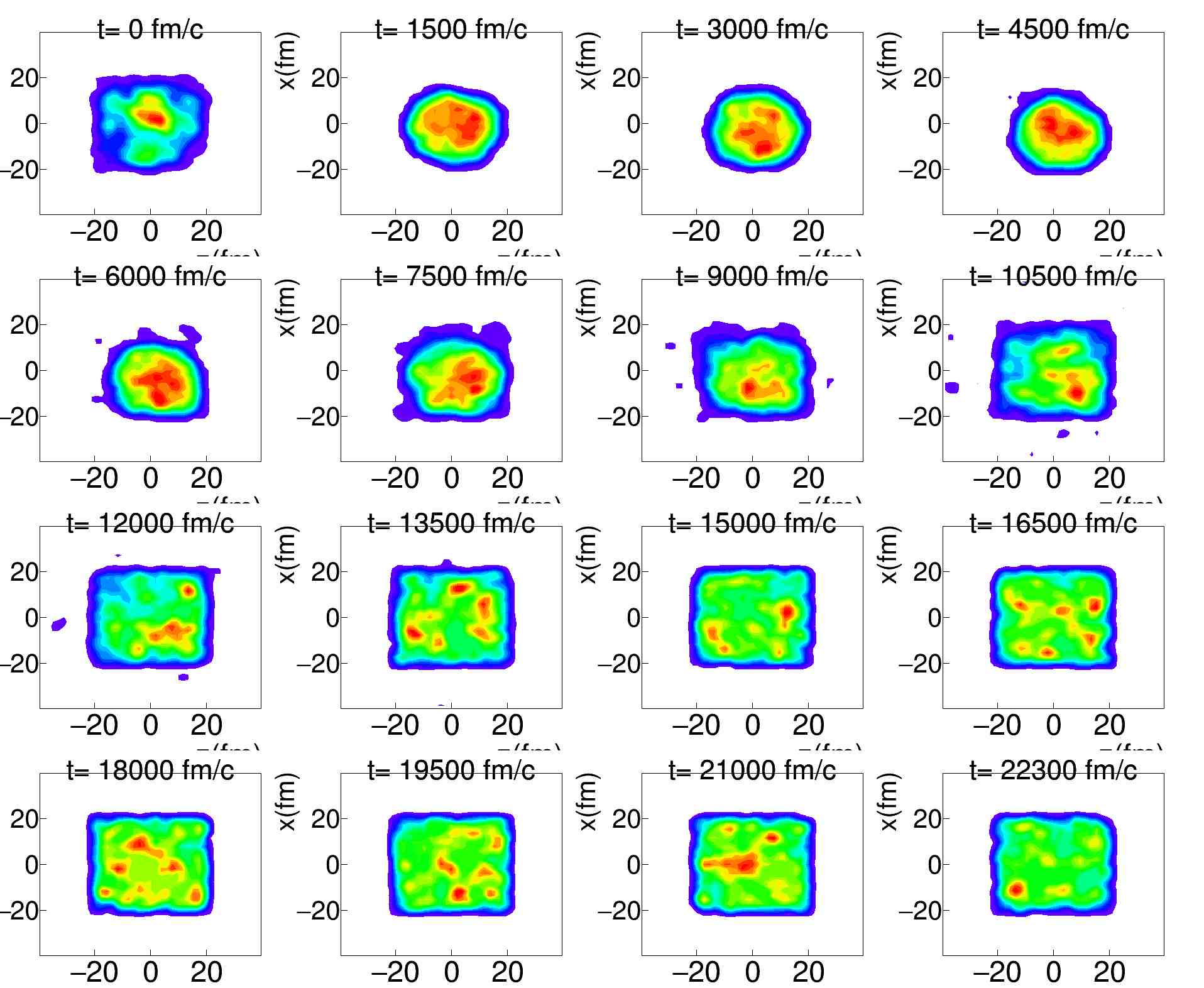}} 
\end{minipage}
\caption{Evolution of the system $^{140}$Ni+$^{460}$U in the nucleon bath with 10\% proton concentration calculated using the CoMD code at beam energy $0.01~MeV/nucleon$ and Coulomb interaction cutoff at 10, 7, 5, 4, 3 and 2 $~fm$ (from top to bottom and left to right), corresponding to nucleon bath densities ${\rho_{0}/100}$, ${\rho_{0}/50}$, ${\rho_{0}/30}$, ${\rho_{0}/17}$, ${\rho_{0}/10}$, and ${\rho_{0}/5}$, respectively. Incompressibility is $K_0 = 254~MeV$  and a soft density dependence of symmetry energy were used. The initial distance is $25~fm$. On the time scale up to 25000 $fm/c$ the resulting nucleus appears to dissolve in the nucleon bath with density above ${\rho_{0}/30}$, even sooner with increasing density.}
\end{figure}

\begin{figure}[h]
\begin{minipage}{\textwidth}
\centering
\fbox{\includegraphics[width=8.6cm]{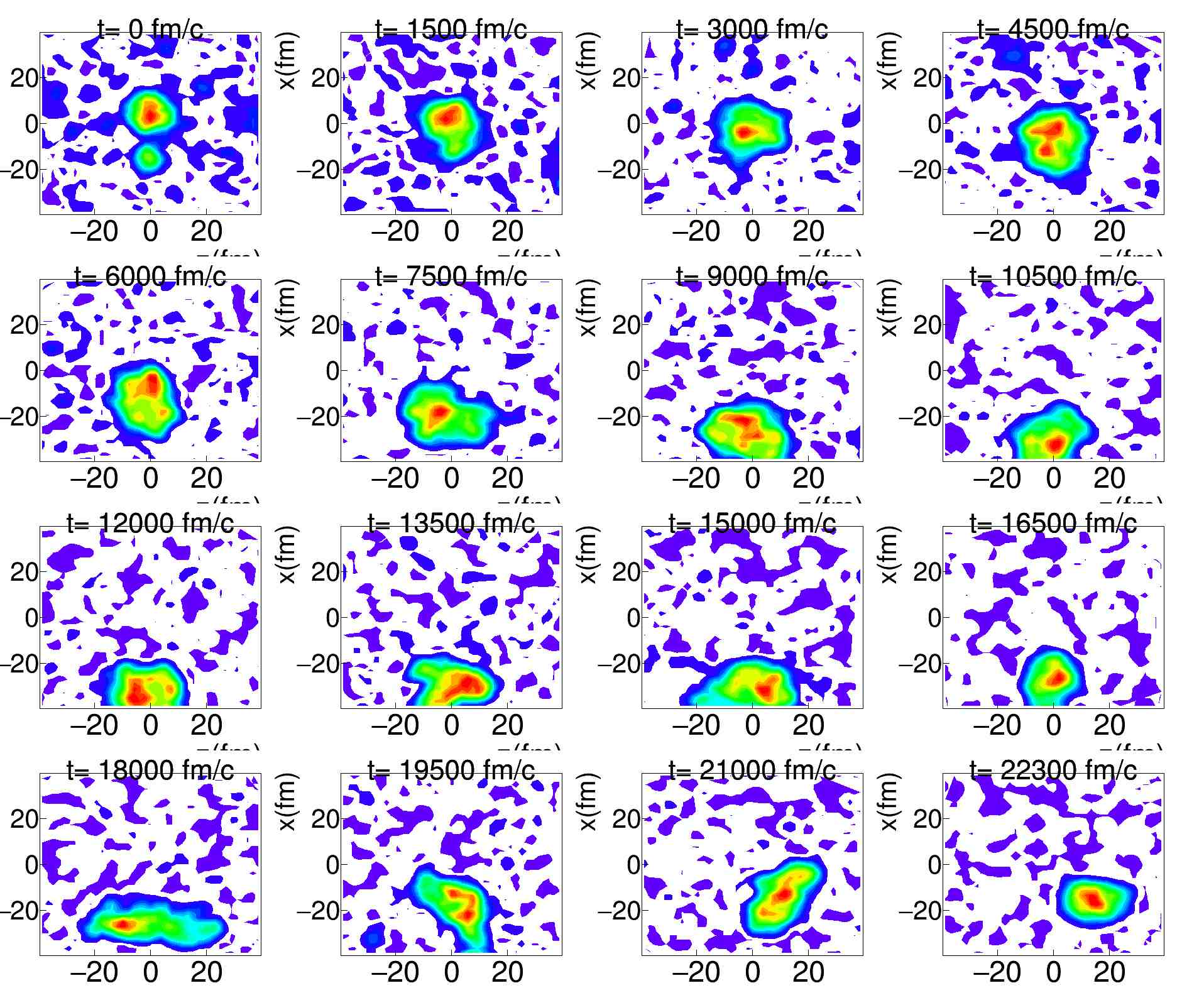}} 
\fbox{\includegraphics[width=8.6cm]{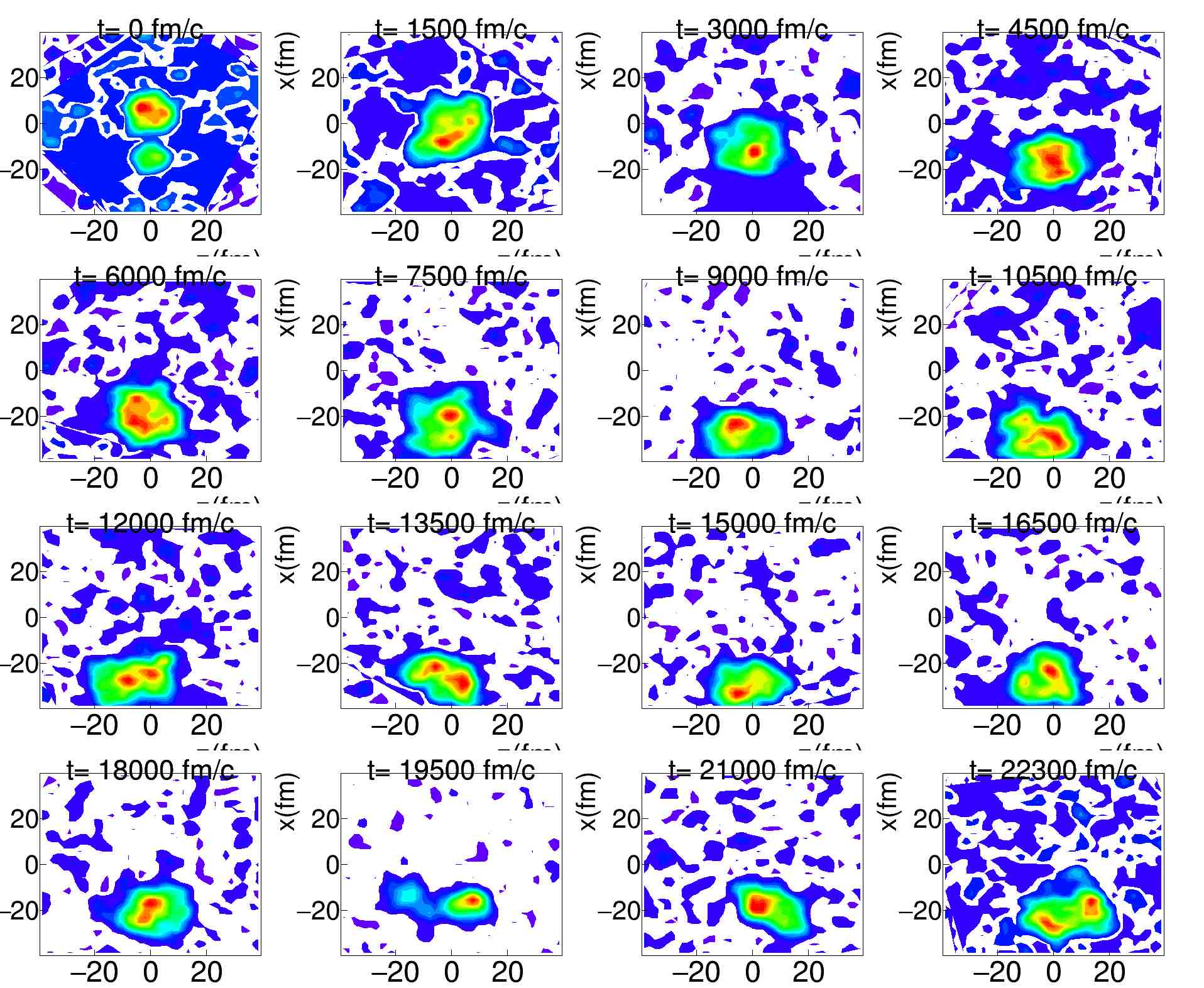}} 
\fbox{\includegraphics[width=8.6cm]{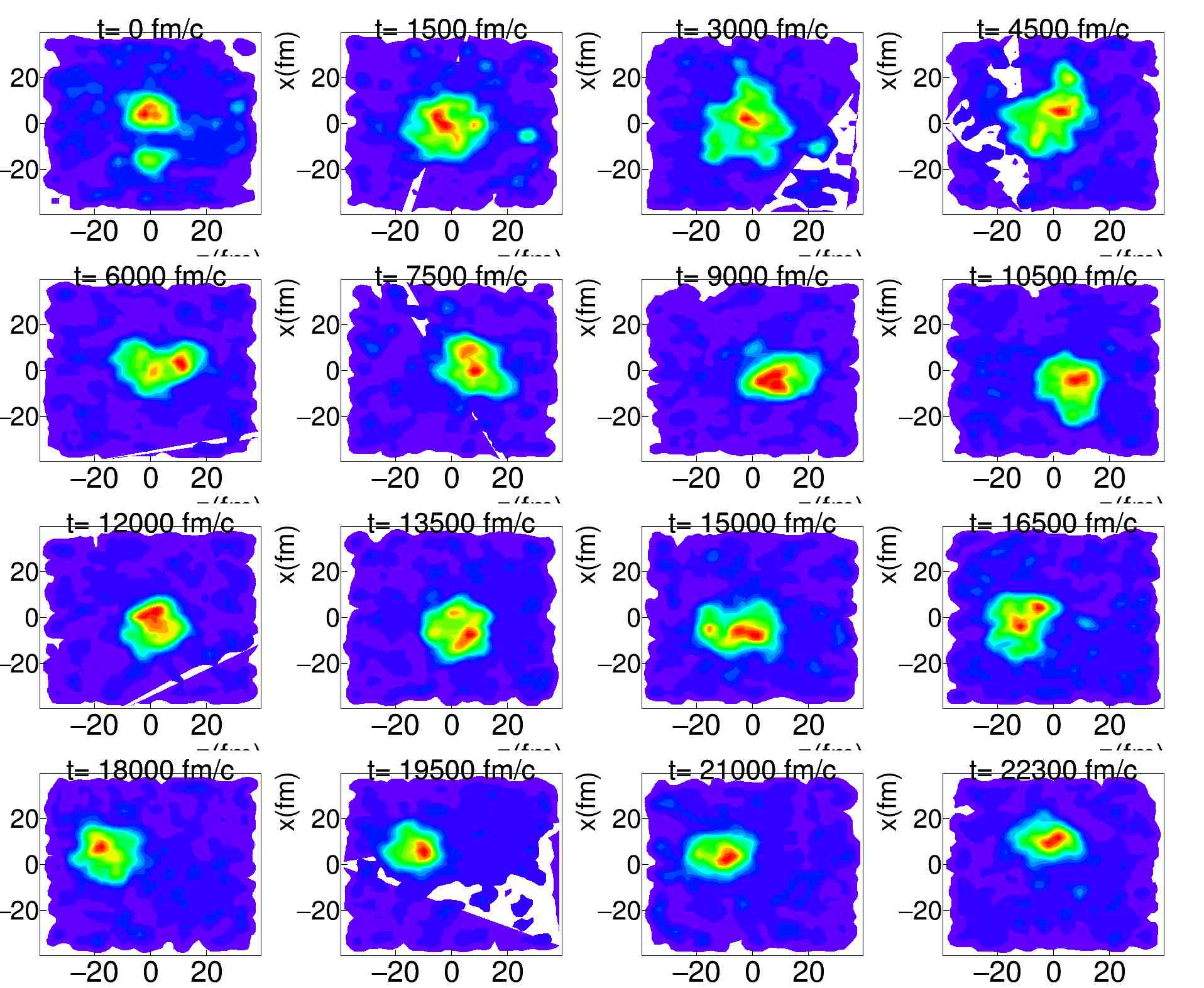}} 
\fbox{\includegraphics[width=8.6cm]{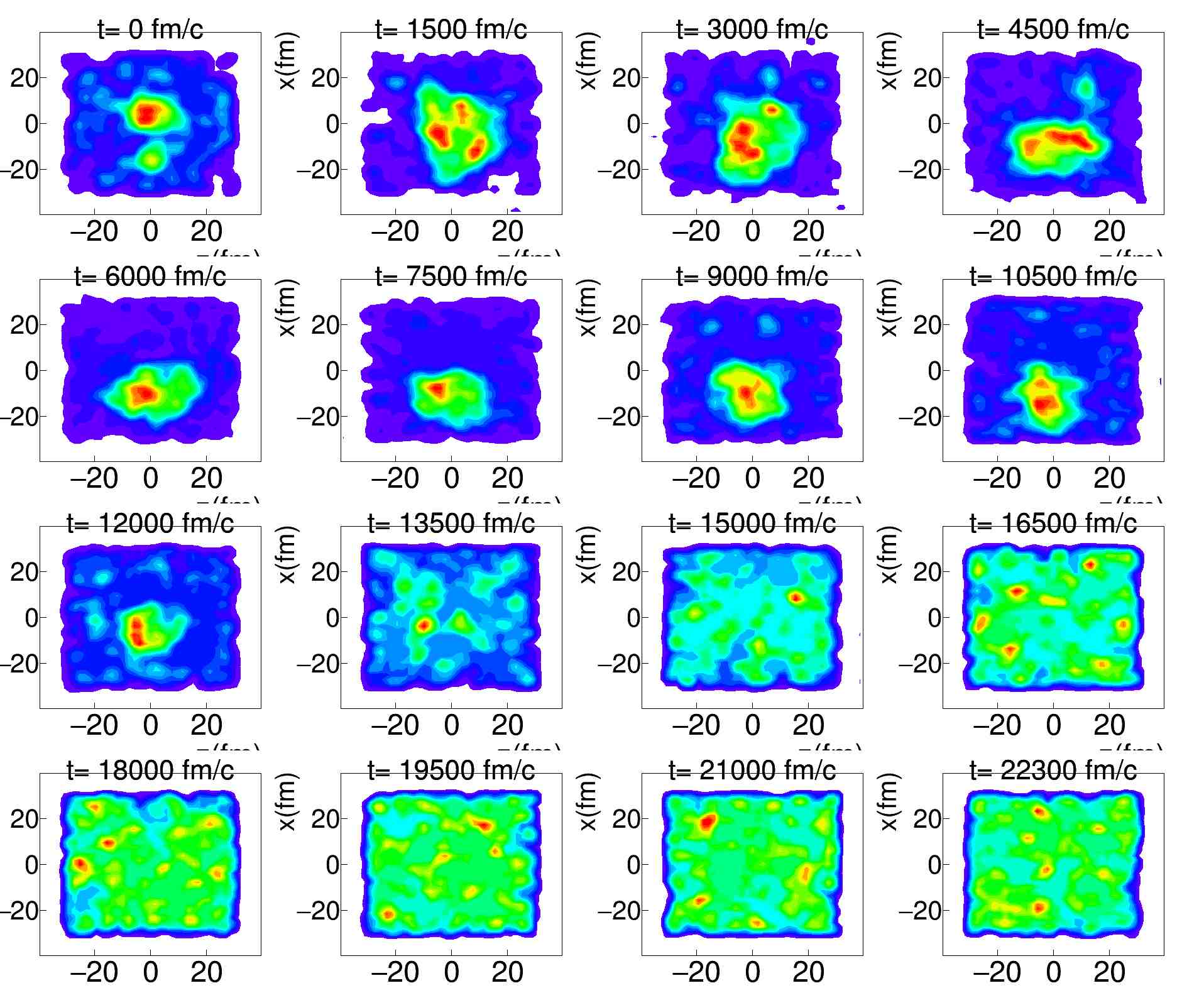}} 
\fbox{\includegraphics[width=8.6cm]{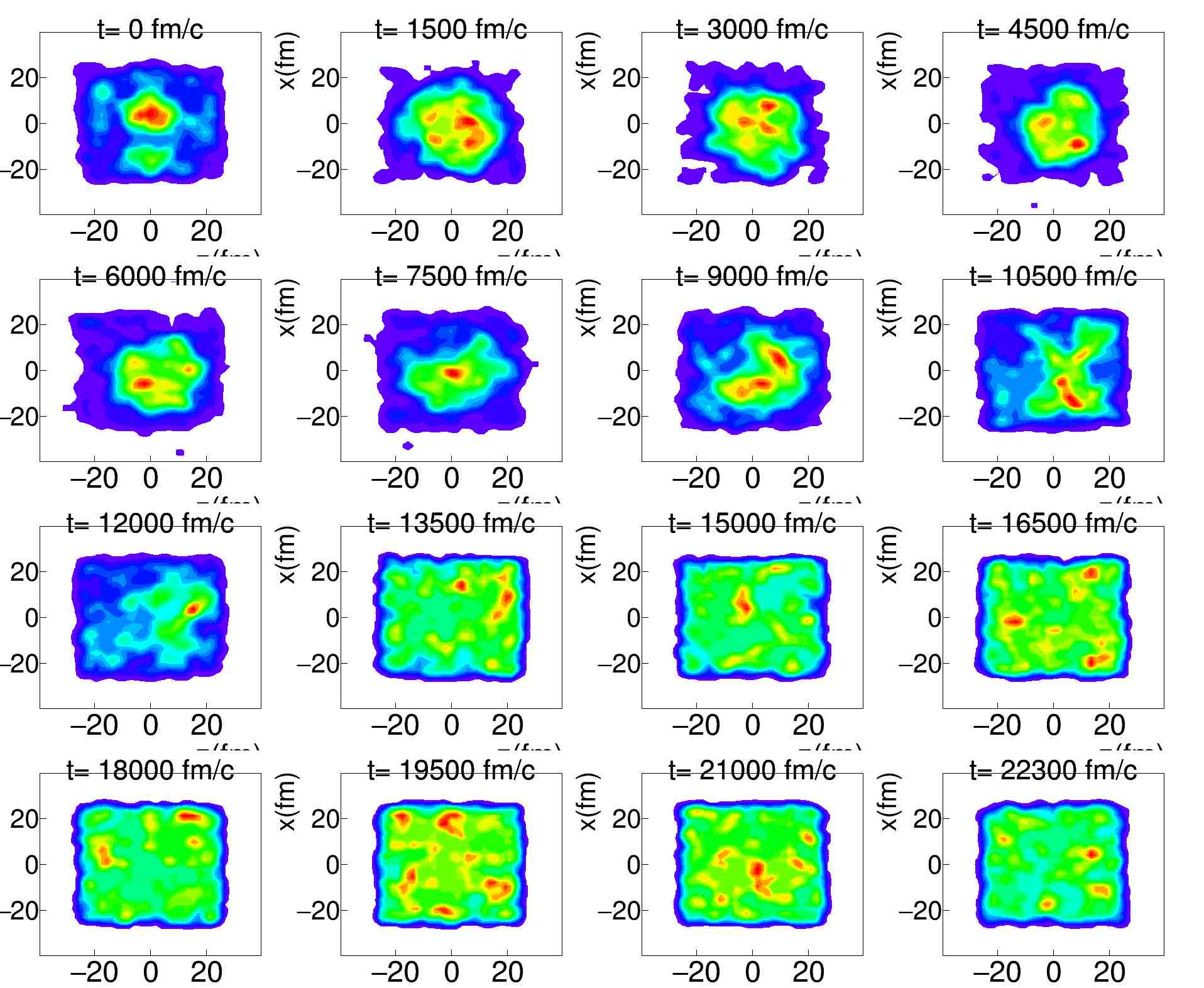}} 
\fbox{\includegraphics[width=8.6cm]{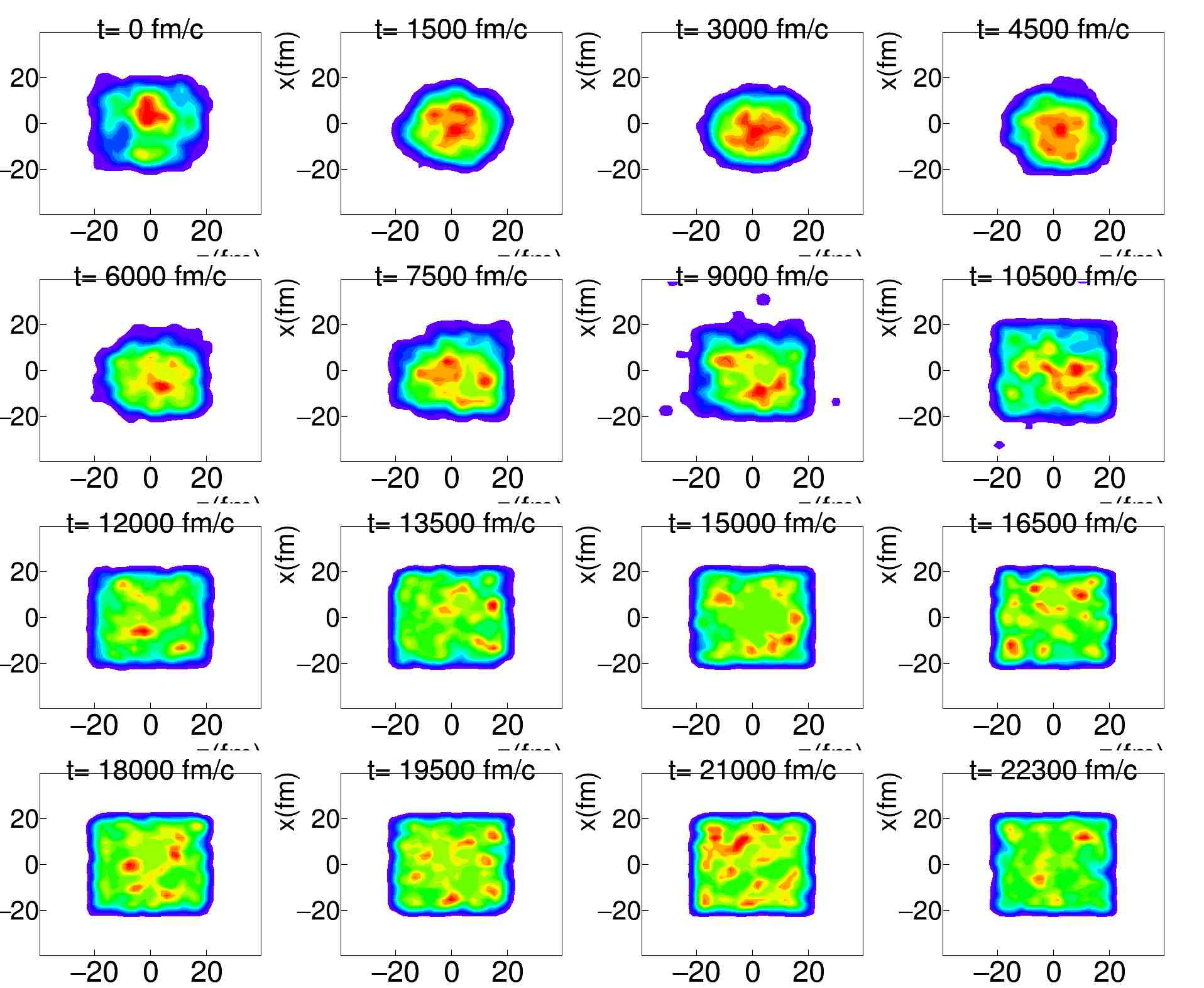}} 
\end{minipage}
\caption{Evolution of the system $^{140}$Ni+$^{460}$U in the nucleon bath with 10\% proton concentration calculated using the CoMD code at beam energy $0.01~MeV/nucleon$ and Coulomb interaction cutoff at 10, 7, 5, 4, 3 and 2 $~fm$ (from top to bottom and left to right), corresponding to nucleon bath densities ${\rho_{0}/100}$, ${\rho_{0}/50}$, ${\rho_{0}/30}$, ${\rho_{0}/17}$, ${\rho_{0}/10}$, and ${\rho_{0}/5}$, respectively. Incompressibility is $K_0 = 254~MeV$  and a medium density dependence of symmetry energy were used. The initial distance is $25~fm$. On the time scale up to 25000 $fm/c$ the resulting nucleus appears to dissolve in the nucleon bath with density above ${\rho_{0}/30}$, even sooner with increasing density.}
\end{figure}

\begin{figure}[h]
\begin{minipage}{\textwidth}
\centering
\fbox{\includegraphics[width=8.6cm]{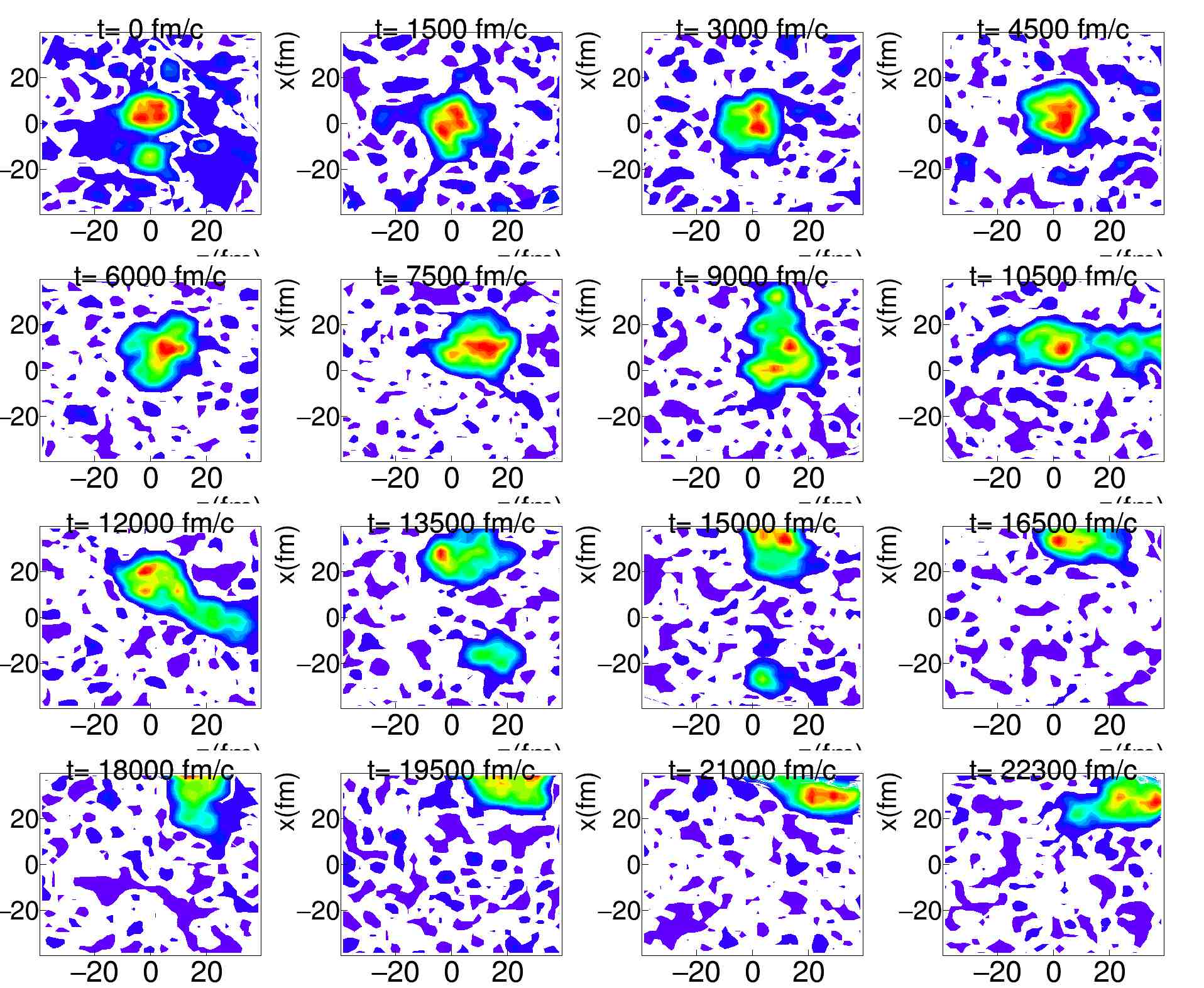}} 
\fbox{\includegraphics[width=8.6cm]{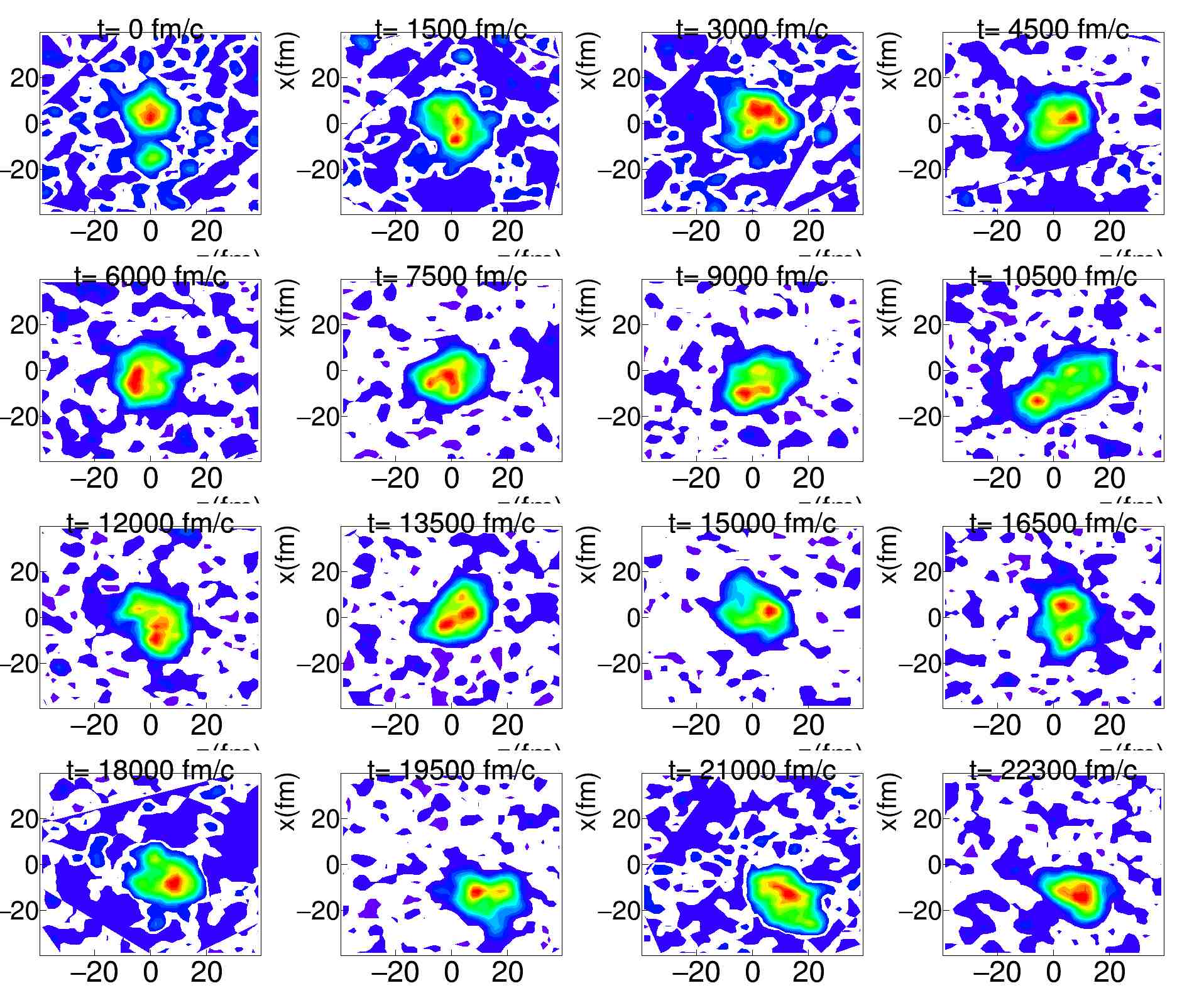}} 
\fbox{\includegraphics[width=8.6cm]{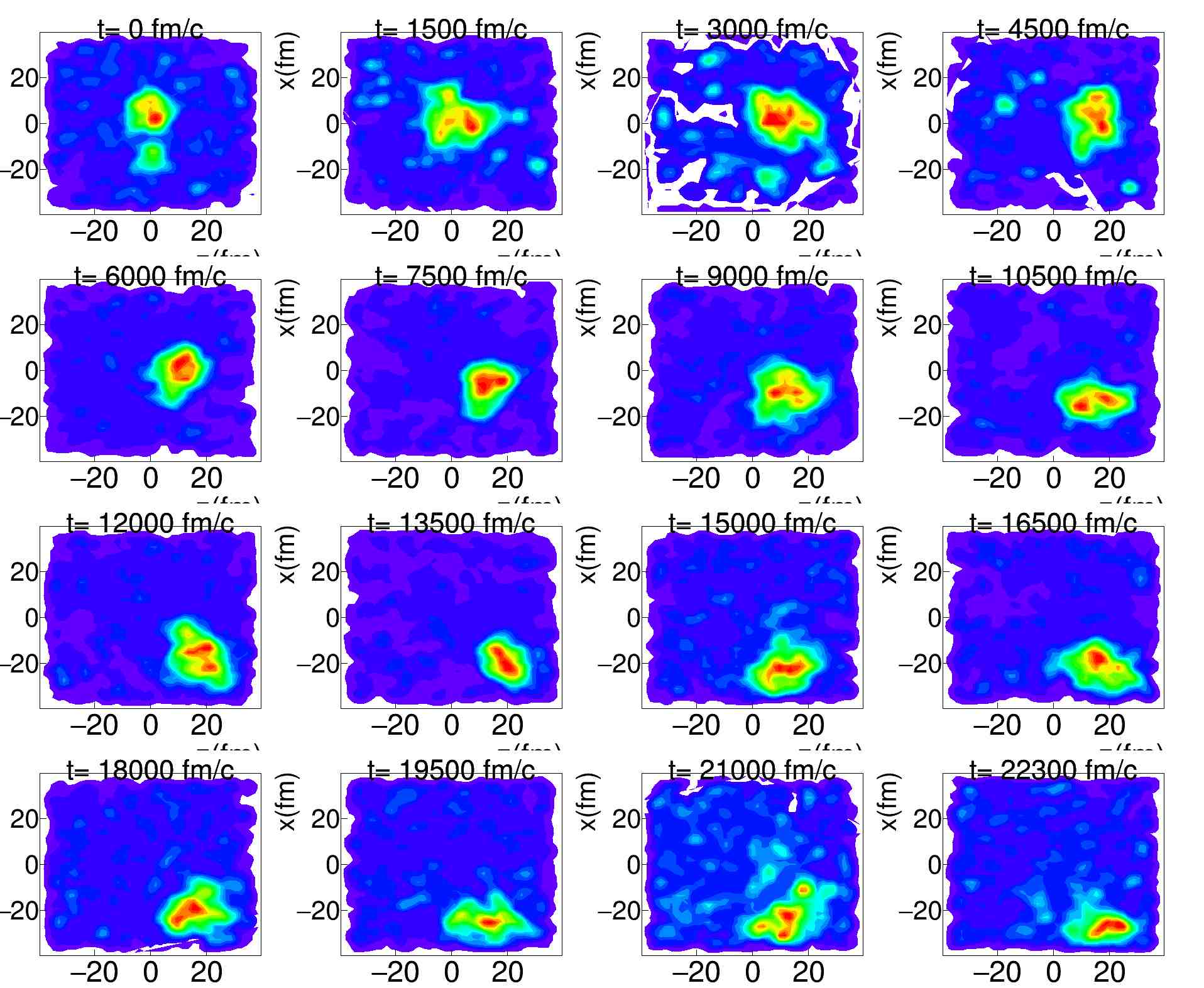}} 
\fbox{\includegraphics[width=8.6cm]{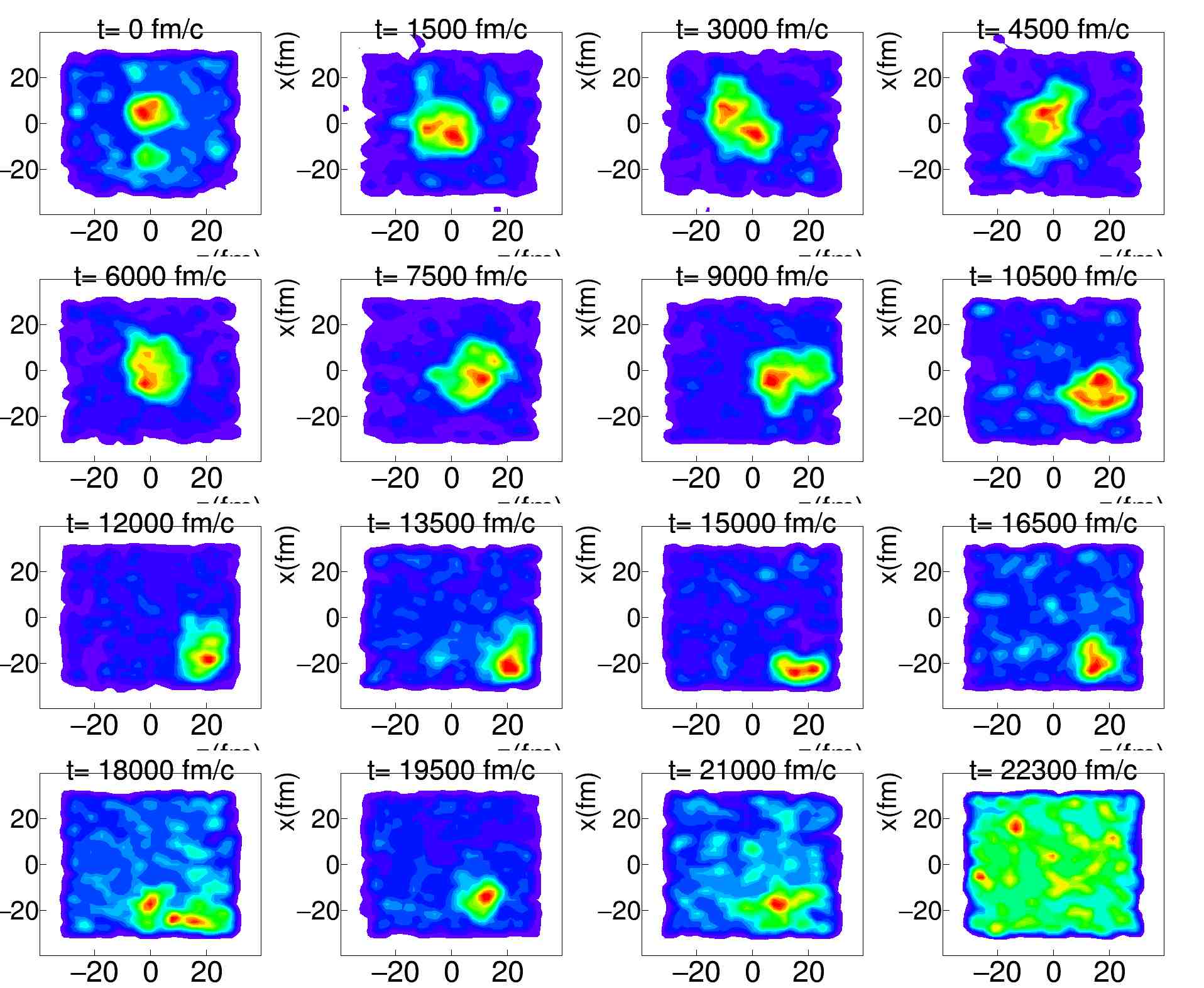}} 
\fbox{\includegraphics[width=8.6cm]{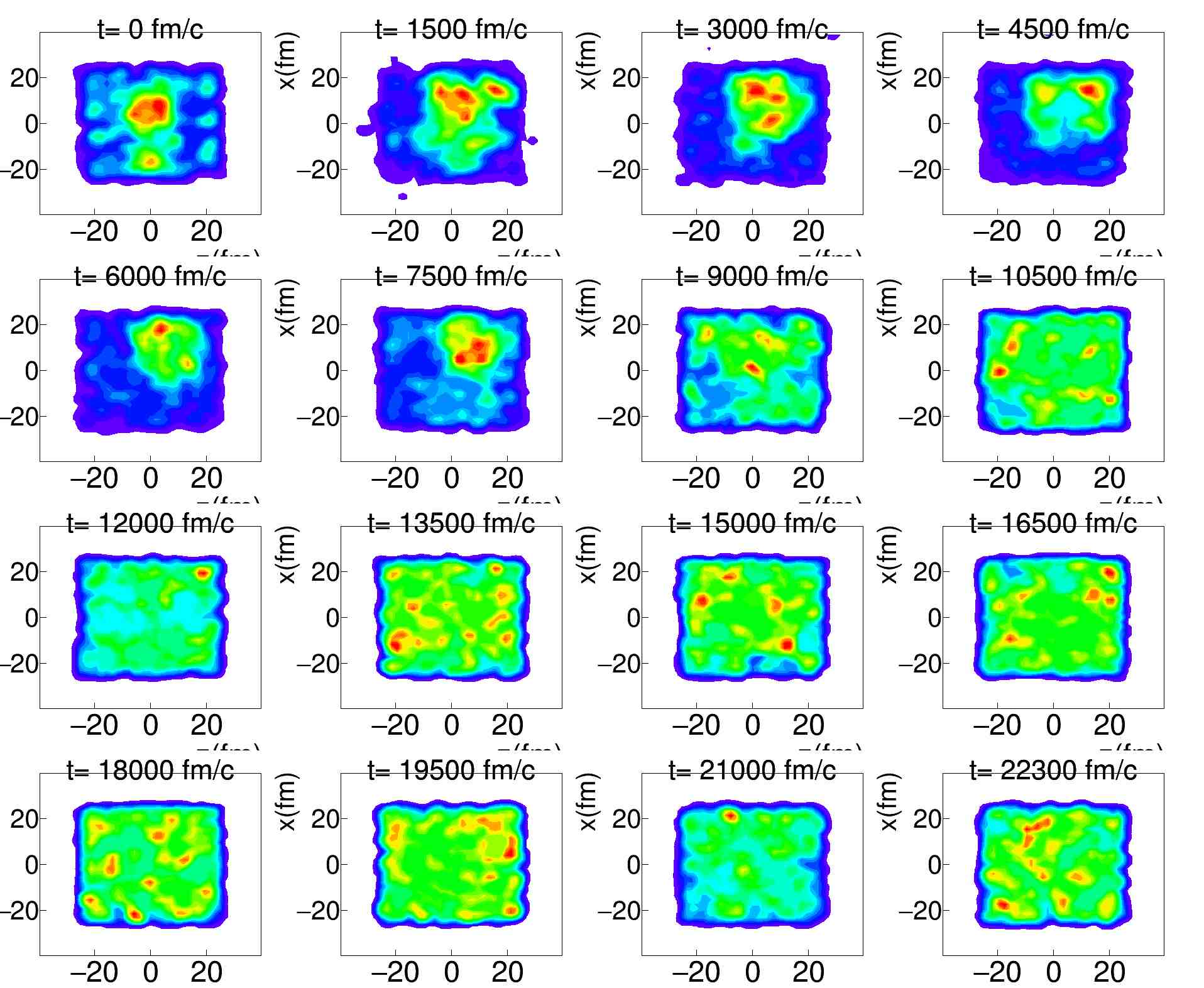}} 
\fbox{\includegraphics[width=8.6cm]{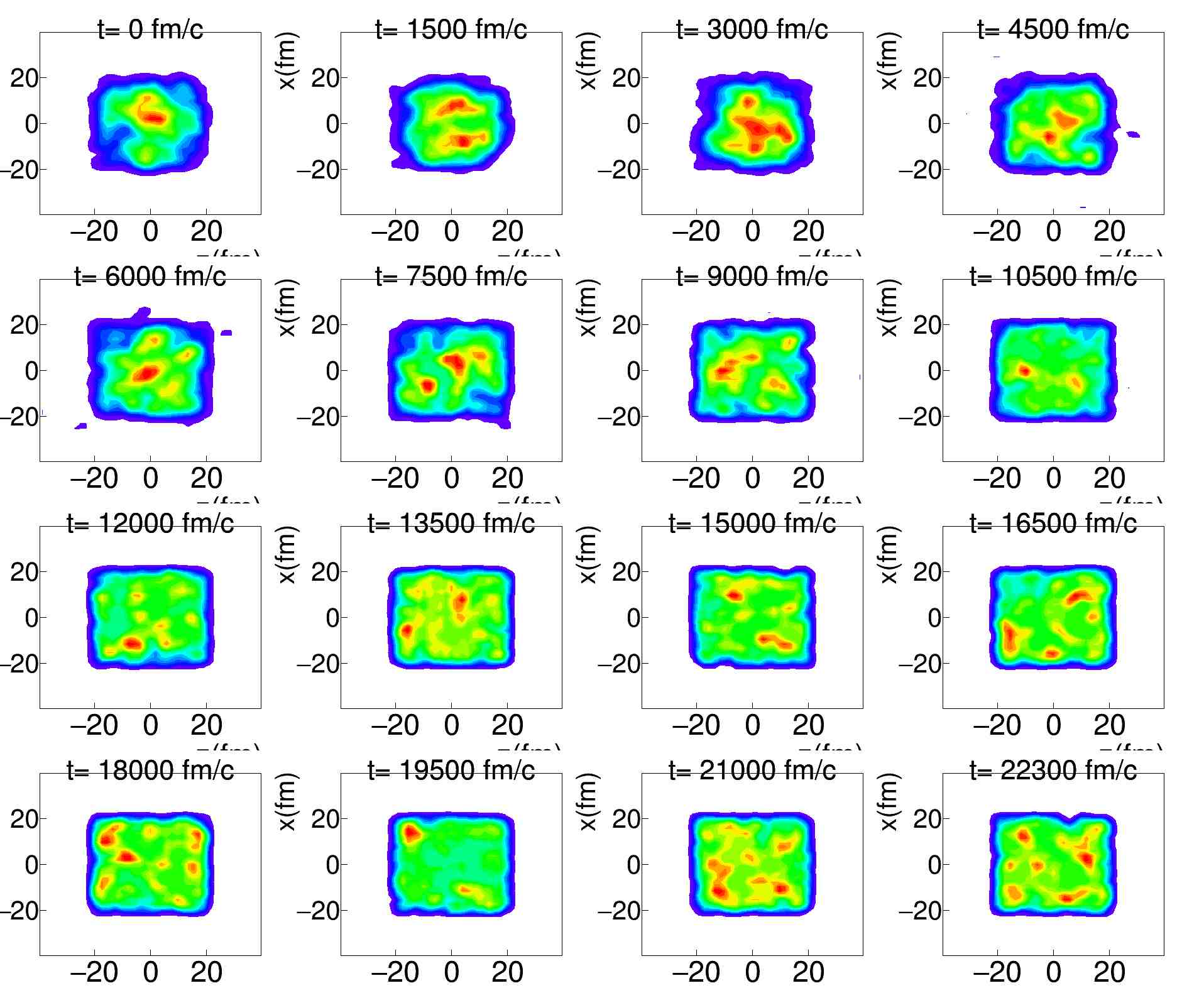}} 
\end{minipage}
\caption{Evolution of the system $^{140}$Ni+$^{460}$U in the nucleon bath with 10\% proton concentration calculated using the CoMD code at beam energy $0.01~MeV/nucleon$ and Coulomb interaction cutoff at 10, 7, 5, 4, 3 and 2 $~fm$ (from top to bottom and left to right), corresponding to nucleon bath densities ${\rho_{0}/100}$, ${\rho_{0}/50}$, ${\rho_{0}/30}$, ${\rho_{0}/17}$, ${\rho_{0}/10}$, and ${\rho_{0}/5}$, respectively. Incompressibility is $K_0 = 254~MeV$  and a stiff density dependence of symmetry energy were used. The initial distance is $25~fm$. On the time scale up to 25000 $fm/c$ the resulting nucleus appears to dissolve in the nucleon bath with density above ${\rho_{0}/30}$, even sooner with increasing density. At lowest density the fission appears to take over, thus suggesting a maximum of life time between densities ${\rho_{0}/50}$ and ${\rho_{0}/30}$.}
\end{figure}

\section{Conclusions}
The synthesis of hyper-heavy elements was investigated under conditions simulating a neutron star environment. The Constrained Molecular Dynamics (CoMD) approach was used to simulate the low energy collisions of extremely n-rich nuclei. A new type of the fusion barrier due to a "neutron wind" has been observed, when the effect of a neutron star environment (screening of Coulomb interaction) was introduced implicitly. When introducing also a nucleonic background of surrounding nuclei, fusion becomes possible down to temperatures 10$^{8}$ K and synthesis of extremely heavy and n-rich nuclei appears feasible. 

Specifically, in this work, the evolution of a hyper-heavy system consisting of $^{140}$Ni+$^{460}$U in a nucleon bath with 10\% proton concentration was calculated using the CoMD code at beam energy from $10~keV$ to $1~MeV$. Coulomb interaction screening was introduced by a cutoff at 10, 7, 5, 4, 3 and 2 $fm$, corresponding to nucleon bath densities ${\rho_{0}/100}$, ${\rho_{0}/50}$, ${\rho_{0}/30}$, ${\rho_{0}/17}$, ${\rho_{0}/10}$, and ${\rho_{0}/5}$, respectively. An EoS with incompressibility $K_0 = 254~MeV$ was used. 

Using a soft, medium and stiff density dependence of symmetry energy, on the time scale up to 25000 $fm/c$, the resulting nucleus appeared to dissolve in the nucleon bath with density above ${\rho_{0}/30}$, even sooner with increasing density. In addition, for a stiff density dependence of symmetry energy and at lowest density, the fission appears to take over, thus suggesting a maximum of lifetime between densities ${\rho_{0}/50}$ and ${\rho_{0}/30}$, sufficient to support a fusion cascade.

We suggest that this possible existence of hyper-heavy nuclei in a neutron star environment could provide an extra coherent neutrino scattering. This behavior could be considered as an another crucial process, adding new information to the suggested models on neutron star cooling rate. 
On the other hand, local events of fusion cascade leading to the production of hyperheavy nuclei can lead to energy release due to minimization of surface energy, that, in turn may lead to an additional mechanism of X-ray bursts. Alternatively, due to the local density profile modification deeper within the neutron star, gravitational wave signals may result from a violation of rotational symmetry~\cite{Meisel}.

\section*{Ackwnoledgments}
This work is supported by the Grant Agency of Czech Republic (GACR Contr.No. 21-24281S). 
The simulations were performed at the Supercomputing facility of Czech Technical University in Prague.

\end{document}